\documentclass[prl,twocolumn,superscriptaddress,aps,footinbib]{revtex4-2}
\usepackage{ulem}
\usepackage{bbm}
\usepackage{color}
\usepackage{graphicx}
\usepackage{footmisc}

\usepackage{amsmath}
\usepackage{hyperref}
\usepackage{comment}
\usepackage{siunitx} 
\usepackage{verbatim}

\usepackage{amssymb}

 \usepackage[caption=false]{subfig}

\DeclareMathOperator{\osc}{osc}

\definecolor{mygreen}{rgb}{0,0.5,0}
\definecolor{mybrown}{rgb}{0.6, 0.4, 0.2}
\definecolor{myyellow}{rgb}{0.99,0.75,0.1}
\definecolor{myred}{rgb}{0.75,0,0}
\definecolor{myblue}{rgb}{0,0,0.75}
\definecolor{mymagenta}{cmyk}{0,1,0,0.12}
\definecolor{mycyan}{cmyk}{1,0,0,0.12}
\definecolor{myorange}{rgb}{1,0.5,0}
\definecolor{myviolet}{rgb}{0.5,0.0,0.75}

\newcommand{\commentout}[1]{}

\newcommand{\modeOne}{\alpha}
\newcommand{\modeTwo}{\beta}
\newcommand{\modeThree}{\kappa}
\newcommand{\modeFour}{\mu}

\newcommand{\be}{\begin{equation}}
\newcommand{\ee}{\end{equation}}
\newcommand{\bea}{\begin{eqnarray}}
\newcommand{\eea}{\end{eqnarray}}

\newcommand{\nne}{\nonumber \\ & = &}
\newcommand{\nnequiv}{\nonumber \\ & \equiv &}

\newcommand{\nnpropto}{\nonumber \\ & \propto &}

\newcommand{\nnt}{\nonumber \\ & & \times}

\newcommand{\ii}{\textbf{i}}
\renewcommand{\ii}{i}

\newcommand{\ket}[1]{\left|#1\right>}
\newcommand{\bra}[1]{\left<#1\right|}
\newcommand{\hEp}{\hat {E}^{(+)}}
\newcommand{\hEm}{\hat {E}^{(-)}}
\newcommand{\ha}{\hat{a}}
\newcommand{\had}{\hat a^{\dagger}}
\newcommand{\om}[1]{\omega_{#1}}

\graphicspath{{Images/}}

\begin{document}

\newcommand{\mytitle}{Non Degenerate HOM or Ghosh Mandel Effect in Time Domain}
\renewcommand{\mytitle}{Autoheterodyne characterization of narrow-band photon pairs}

\title{\mytitle}

\newcommand{\ICFO}{ICFO - Institut de Ci\`encies Fot\`oniques, The Barcelona Institute of Science and Technology, 08860 Castelldefels (Barcelona), Spain}
\newcommand{\ICREA}{ICREA - Instituci\'{o} Catalana de Recerca i Estudis Avan{\c{c}}ats, 08010 Barcelona, Spain}
\newcommand{\JUAddress}{Institute of Physics, Jagiellonian University in Krak\'ow, \L{}ojasiewicza 11, 30-348 Krak\'ow, Poland}

\author{Vindhiya Prakash}\email[]{vindhiya.prakash@icfo.eu}
\affiliation{\ICFO}

\author{Aleksandra Sierant}
\affiliation{\JUAddress}

\author{Morgan W. Mitchell}\email[]{morgan.mitchell@icfo.es}
\affiliation{\ICFO}
\affiliation{\ICREA}



\begin{abstract}
We describe a  technique to measure photon pair joint spectra by detecting the time-correlation beat note when non-degenerate photon pairs interfere at a beamsplitter.  The technique implements a temporal analog of the Ghosh-Mandel effect with one photon counter and a time-resolved Hong-Ou-Mandel interference with two. It is  well suited to characterize pairs of photons, each of which can interact with a single atomic species, as required to study recently predicted photon-photon interaction in sub-wavelength atomic arrays.  With this technique, we characterize photon pairs from cavity-enhanced parametric downconversion with a bandwidth $\approx \SI{5}{\mega\hertz}$ and frequency separation of $\sim \SI{200}{\mega\hertz}$ near the D$_1$ line of atomic Rb. 
\end{abstract}

\maketitle

Spontaneous parametric downconversion (SPDC) is a ubiquitous technique in photonic quantum technology, where it is used to generate entangled photons with tailored spectral, spatial, and polarization properties \cite{couteau2019SPDC}. The frequency correlations of SPDC photon pairs, including time-frequency entanglement, are of particular importance.  In some applications these correlations are used to encode quantum information 
\cite{ReimerNP2019, PfisterJPB2019}.
In others, the frequency correlations are an undesired side-channel that reduces nonclassical interference \cite{gerrits2015hom}. These correlations are revealed through analysis of the joint spectral amplitude (JSA) or joint spectral intensity (JSI) of the downconverted photon pair. 

In broadband SPDC applications, it is possible to directly measure the JSI using monochromators or other passive filters \cite{Kim:05,zielnicki2018jsi}. Techniques such as Fourier transform spectroscopy using Mach Zehnder interferometers \cite{Wasilewski:06} and temporal magnification of photons with a time lens  \cite{mittal2018temporallens} have also been used.
Non-classical interference can also be a tool to characterize non-classical frequency correlations; the Hong-Ou-Mandel (HOM) \cite{hong1987hom} interference visibility has been used to characterize broadband photon pairs from a single source \cite{gerrits2015hom, Gerrits2011hom, zielnicki2018jsi} and from different sources \cite{wang2019research,Thiel:20}.

Photon pairs with $\sim \SI{}{\mega\hertz}$ bandwidths are important for applications in quantum information \cite{slattery2019background} where material systems such as atoms or ions serve as storage or processing units
\cite{DistanteNC2017, HughRiedmatten2017a}.  For example, sub-wavelength arrays of neutral atoms support sub-radiant states \cite{AsenjoGarciaPRX2017} that can exhibit topological protection \cite{PerczelPRL2017} and unprecedented optical properties \cite{RuiN2020}. Strong photon-photon interactions \cite{MassonPRR2020} that could be harnessed for photonic quantum-information processing, and photonic bound states \cite{DeutschPRL1992, KePRR2020} are predicted in such arrays. Exploring this physics motivates non-classical light sources in which both photons are resonant to an atomic transition \cite{prakash2019narrowband}. {Applications in quantum networking, e.g. entanglement-swapping with memory-compatible photons, will require pure, indistinguishable, narrowband photons \cite{Monteiro_14}.} 

Such narrow-band photon pairs are not easily measured by passive frequency-domain techniques,  because of the very high optical frequency resolution it would require. In such a two-photon Fock state, first-order interference vanishes, producing no observable beat note \cite{Note1}.  One alternative is stimulated parametric downconversion \cite{SipeSET}, in which laser photons are  used to seed the downconversion and map the difference frequencies generated vs those suppressed \cite{Jeong:20, prakash2019narrowband}. This technique has potential for use in tomography of the JSA \cite{Jizan:16}, but requires an additional well-characterized laser source and careful matching of spatial modes.  

Here we present a simpler and more efficient alternative, a time-domain characterization of the {two-photon state} using non-classical interference.  
The {JSI} is a two-dimensional function, while the HOM interference visibility is a scalar observable. {Thus a characterisation of the JSI even along a single dimension requires many HOM visibility measurements under changing experimental conditions, such as a changing path length \cite{OuPRL1988,Jin_18}.} In contrast, the Ghosh and Mandel experiment \cite{GhoshPRL1987} (GM), which measured the spatial interference pattern produced by photons of unequal momentum, showed how a correlation spectrum can be acquired with a single experimental condition.  This motivates us to look for techniques that give more direct and more efficient access to the frequency correlations of interest. Our proposals for narrow-band photon pair characterization, are extensions of the HOM and GM interference effects.

\begin{figure}[t]
\centering
\includegraphics[width=0.99\columnwidth]{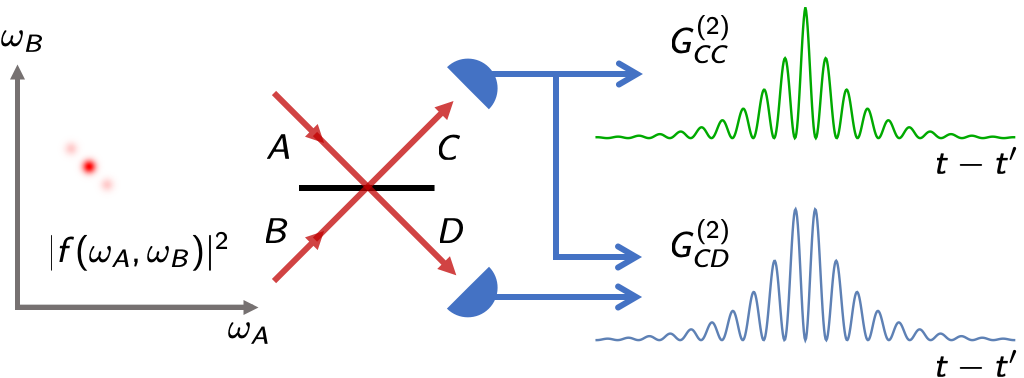} 
\caption[justification=justified]{Principle of the method.  Two photons, one in mode A and one in mode B, with joint spectral amplitude $f(\omega_A, \omega_B)$, illustrated in left graph ( color density indicates square magnitude), meet at a 50:50 BS and are detected in modes C and D. Single-photon-sensitive detectors register the photon arrival times.  The correlation functions $G^{(2)}_{C,C}(t,t')$, $G^{(2)}_{D,D}(t,t')$ and $G^{(2)}_{C,D}(t,t')$ oscillate with $t-t'$, revealing the distribution of $\omega_A-\omega_B$. } 
\label{fig:principle}
\end{figure}

The principle of the method is illustrated in \autoref{fig:principle}. Narrow-band photon pairs with the photons of each pair matched in polarization and spatial profile but with unequal frequency, are injected, one into port $A$ and the other into port $B$ of a 50/50 beamsplitter (BS).  Single-photon-sensitive detectors register photons leaving the BS by ports $C$ and $D$, and time-tagging electronics record the arrival times.  Many events are accumulated, and the {second order} correlation functions $G^{(2)}_{\modeThree,\modeFour}(t,t')$, $\modeThree, \modeFour \in \{C,D\}$ are calculated. 
Information about the JSA can then be inferred from the $G^{(2)}$ functions. We refer to this method as \textit{auto-heterodyne characterization} (AHC), not to be confused with the single-photon self-heterodyne technique \cite{Okawa:17}.

 When photons of different frequency meet at a BS, their arrival-time distribution becomes modulated at the difference of their frequencies.  This can be understood as follows: a detection at $C$ and $D$ with zero time delay, i.e. $t = t'$, can happen by two channels in configuration space: either reflection of both photons or transmission of both photons. The amplitudes for these channels sum to zero, due to phase factors in the transmission and reflection processes. The resulting vanishing of $G^{(2)}_{C,D}(t,t)$ and the corresponding increase of $G^{(2)}_{C,C}(t,t)$ and $G^{(2)}_{D,D}(t,t)$ is the well-known HOM effect \cite{hong1987hom}.  For unequal detection times, one must also consider the phase factors $\exp[-i \omega_A t - i \omega_B t']$  and $\exp[-i \omega_A t' - i \omega_B t]$ that apply to the two-reflection and two-transmission channels, respectively.   The relative phase $(\omega_A-\omega_B) (t - t')$ between the channels then induces an oscillation of the $G^{(2)}$ correlations at the difference frequency $\omega_A-\omega_B$.  We refer to this non-classical interference between distinguishable photons as the non-degenerate HOM effect.   Small frequency differences between the photons manifest as long-period oscillations in the relative arrival time distribution, which are technologically convenient to detect.

The effect can be easily calculated. For a two-photon input state $\ket{\psi}$ with JSA {$f(\om{A},\om{B})$},
\be
\label{eq:PsiDef}
\ket{\psi} = \int d\om{A}\, d\om{B}\, f(\om{A},\om{B}) \had_A(\om{A}) \had_B(\om{B}) \ket{0}.
\ee
The {un-normalised} correlation functions are
$G^{(2)}_{\modeOne,\modeTwo}(t,t') \equiv  |\langle 0 | \hEp_\modeOne(t) \hEp_\modeTwo(t')  | \psi \rangle|^2$,  $\modeOne,\modeTwo \in \{A,B,C,D\}$,
 where the field operators are $\hEp_\modeOne(t) \propto \int d\omega \ha_\modeOne(\omega) \exp[{-i\omega t}]$ and $\ha_\modeOne$ is an annihilation operator. Due to the beamsplitter, the 
output fields are $\hEp_{C/D}(t)  \propto  \hEp_A(t)  \pm \hEp_B(t)$. A straightforward calculation finds

\bea
\label{eq:G2HOM}
G^{(2)}_{\modeThree,\modeFour}(t ,t')  
&\propto& \left| \int d^2\omega f(\vec{\omega})  e^{-\ii \frac{ \omega_+ t_+ }{2}}
\osc({\omega_- t_- }/{2}) \right|^2, \hspace{6mm}
\eea
where $\omega_\pm \equiv \omega_A \pm \omega_B$, $t_\pm \equiv t \pm t'$ and the integral is taken over $\omega_A, \omega_B$ or equivalently over $\omega_+,\omega_-$, with $\vec{\omega}$ being the corresponding parametrization of $f$ and $\osc \theta = \frac{1}{\sqrt{2}}\cos \theta$ if $\modeThree,\modeFour = C,C$ or $D,D$  and $i \sin \theta$ otherwise. This describes a Fourier transform of the JSA along the $\omega_\pm$ coordinates. When the two-photon state is produced by SPDC, and pumped by a broadband pump with field $E_p(t)= \int d\omega_p\alpha(\omega_p)e^{-i \omega_p t}$, $f \propto \int d\om{p}\alpha(\omega_p) \delta(\omega_p - \om{+})g(\om{-},\om{+}) \propto \alpha(\om{+})g(\om{-},\om{+})$. When the variation of the crystal phase matching function over the pump-bandwidth can be neglected, $g$ becomes independent of $\omega_+$ and the JSA factorizes as  $f \propto \alpha(\om{+})g(\om{-})$. The $G^{(2)}$ then gives the sine or cosine power spectrum of $g(\om{-})$ in the $t_-$ dimension and also gives the Fourier transformed spectrum of $\alpha(\om{+})$ via the $t_+$ dimension \footnote{See Supplemental Material for application of AHC to heralded pure state characterisation, calculations relating the power spectral density of the AHC $G^{(2)}$ to the JSA, discussions related to the resolution of AHC and first order interference effects from single photon states.}. 
For a monochromatic pump $\alpha(\om{+}) \rightarrow \delta(\om{+}-\om{p})$, such that  $f \propto \delta(\omega_+ - \om{p})g(\om{-})$, and $G^{(2)}$ depends only on $t_-$. In what follows we study the narrowband, cw-pump case. The use of AHC in the pulsed scenario and to obtain measures such as entanglement entropy, state purity and Schmidt number \cite{BranczykNJP2010} is discussed in the Supplemental Material \cite{Note1}

We note that this technique implements a variant of the GM effect \cite{GhoshPRL1987}.  In GM, photon pairs with unequal transverse momenta $k_s, k_i$ are observed to produce a spatial auto-correlation function $G^{(2)}(x-x')$ that is maximum for $x-x' = 0$ and modulated with momentum $k_s - k_i$.  The temporal modulation of $G^{(2)}_{C,C}$ or $G^{(2)}_{D,D}$, which describes the correlations of photon pairs with unequal frequencies $\omega_A, \omega_B$ in a single output channel, is the temporal analog of GM.  We refer to this as the temporal Ghosh-Mandel effect. {By conservation of probability at the BS, the GM and HOM signals must add to give $G^{(2)}_{A,B}(t,t')$, as illustrated in \autoref{fig:principle}. As a result, the two methods give very similar information about the JSA.}


\begin{figure}[t]
\centering
 \includegraphics[scale=0.3]{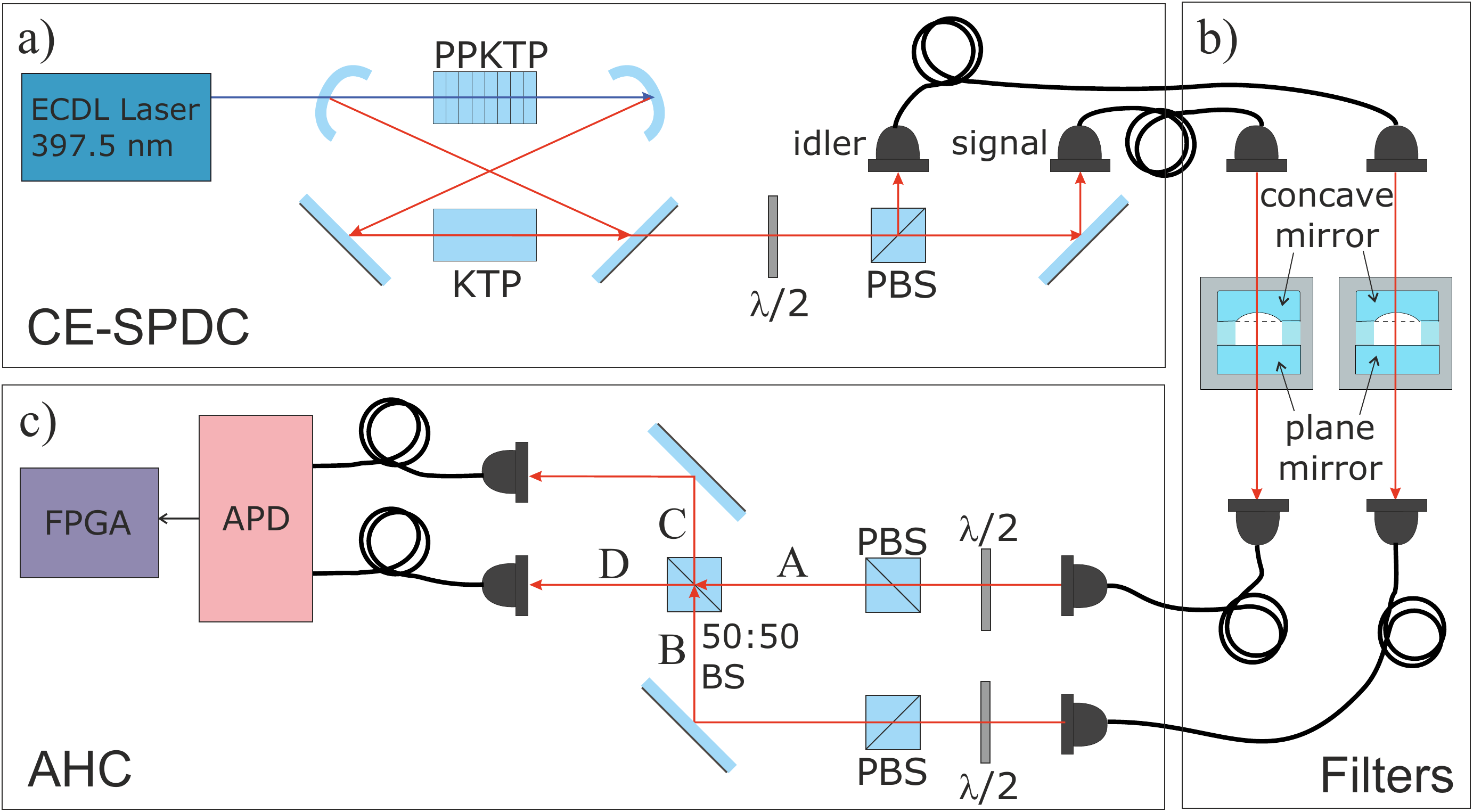} 
\caption{Schematic of setup for generation and characterization of narrow-band photon pairs. (a) CE-SPDC source, consisting of a bow-tie cavity containing an ECDL-pumped SPDC crystal (PPKTP) and a second crystal (KTP). Photon pairs are separated by polarisation. (b) Tuneable FP filters are used to select desired ``teeth'' from the comb of CE-SPDC output modes. (c) Auto-heterodyne characterization. The photons are set to the same polarization, interfered on a beam splitter (BS), detected using avalanche photodiodes (APDs) and time-tagged with an FPGA.  KTP - potassium titanyl phosphate crystal; PPKTP - periodically-poled KTP crystal; PBS - polarizing beam splitter; $\lambda/2$ - half-wave plate; FPGA - field programmable gate array.
}
\label{fig:scheme}
\end{figure}

To demonstrate this experimentally, we use photon pairs from a cavity-enhanced SPDC (CE-SPDC) source pumped by a single-frequency laser. The experimental setup is shown schematically in Fig. \ref{fig:scheme}. 
Photon pairs are produced by a narrow-band CE-SPDC source, described in detail in \cite{prakash2019narrowband}.  A pump laser at \SI{397.5}{\nano\meter} with  full width half-maximum  (FWHM) linewidth $\approx \SI{2}{\mega\hertz}$ pumps a type-II periodically-poled Potassium Titanyl Phosphate (PPKTP) crystal to produce photon pairs of orthogonal linear polarization. The cavity mode structure, with signal and idler free-spectral range (FSR$_s$ and FSR$_i$) $\approx\SI{500}{\mega\hertz}$  and FWHM linewidth  $ \gamma =2\pi\times\SI{7.6}{\mega\hertz}$ for both $H$ (signal - $s$) and $V$ (idler - $i$) polarizations, shapes the output spectrum by the Purcell effect.  Birefringent crystals in the SPDC cavity produce a mismatch of $\Delta$FSR = \SI{3.5}{\mega\hertz} in the FSRs of the signal and idler modes. This reduces the number of modes at the output to 3 clusters of 4 modes each for a downconversion crystal bandwidth of $\SI{150}{\giga\hertz}$ as shown in \cite{prakash2019narrowband}. The contribution of unwanted modes is blocked by a pair of tuneable Fabry-Perot (FP) filters with linewidth $\gamma_{f}=2\pi\times \SI{97}{\mega\hertz}$ and FSR$_{f}=\SI{39}{\giga\hertz}$. 

To perform AHC of the CE-SPDC source, the pump laser, cavity mode and FP mode frequencies are tuned to produce and pass a single signal-idler mode pair within the \SI{794.7}{\nano\meter} D$_1$ line of atomic Rb, with frequency difference $\omega^0_- =\omega^0_s - \omega^0_i$, between the central frequencies. The measurement is performed for two values of  $\omega^0_-/(2\pi)$, $\SI{250}{\mega\hertz}$ and $\SI{165}{\mega\hertz }$. Single-mode fibers and linear polarizers ensure good spatial and polarization matching when the signal and idler photons arrive at the BS via spatial modes $A$ and $B$, respectively.  BS output modes $C$ and $D$ are coupled into single mode fibers leading to avalanche photodiodes (APDs, Perkin Elmer SPCM-AQ4C, quantum efficiency $\approx \SI{50}{\percent}$).  A field programmable gate array (FPGA) records all APD firings with a resolution of \SI{625}{\pico\second} (sampling rate \SI{1.6}{\giga\hertz}).

The two-photon state at the output of a CE-SPDC system when pumped by a cw laser of frequency $\om{p}$ is given by \autoref{eq:PsiDef} with $\omega_{A/B} \rightarrow \omega_{s/i}$, $f\rightarrow f'$ and
\bea
\label{eq:fprime}
 f'(\om{s},\om{i}) &=&  \delta(\om{p}-\om{i}-\om{s}) F(\om{s},\om{i}) 
\nnt {\cal A}(\om{s},\gamma ,\om{s}^0 , \text{FSR}_s )
{\cal A}(\om{i},\gamma ,\om{i}^0 , \text{FSR}_i), \hspace{6mm}
\eea
where $F(\om{s},\om{i})$ is the collinear phase-matching amplitude for the SPDC process and the delta function imposes energy conservation \cite{OBenson3}. For $\nu \in {s,i}$, 
\be
\label{eq:ADef}
 {\cal A}(\omega_\nu,\gamma ,\omega^0_\nu , \text{FSR}_\nu ) = \sum_{m} \frac{\sqrt{\gamma/2 \pi}}{ \frac{\gamma}{2}+\ii\:(\omega^0_{\nu}+m\: \text{FSR}_\nu-\omega_\nu)}
\ee
where the mode index $m$ is summed over positive and negative integers.  In practice, the summations can be truncated to cover only those values for which $F(\omega_s,\omega_i)$ is significant.  

When FP cavities are used to filter the multimode CE-SPDC output, the two-photon JSA after the filters is
\bea
f(\om{s},\om{i}) &=&  f'(\om{s},\om{i}) {\cal A}(\om{s},\gamma_{f} ,\om{f_s}^0 , \text{FSR}_f )
\nnt {\cal A}(\om{i},\gamma_f ,\om{f_i}^0 , \text{FSR}_f ).
\eea
If the FP filters are tuned to the preferred CE-SPDC output modes as described above, the filters' index frequencies coincide with those of the CE-SPDC cavity, i.e., $\omega^0_{f_\nu} = \omega^0_{\nu}$. In this case, the JSA for our FP and CE-SPDC linewidths and FSRs, is well approximated by a Lorentzian each for the signal and idler 
\bea
\label{eq:JSA}
f(\omega_s,\omega_i)&\propto& \delta(\omega_p-\omega_+) \prod_{\nu\in\{s,i\}} \frac{\sqrt{\gamma/2\pi}}{{\gamma}/{2}+\ii\:(\omega^0_{\nu}-\omega_\nu)} 
\nnpropto  \delta(\omega_p - \om{+})\frac{1}{\gamma^2+(\omega^0_{s}- \omega_i^0-\omega_-)^2} \nnequiv  \delta(\omega_p - \om{+}) g(\omega_-).
\eea

When inserted into \autoref{eq:G2HOM}, we find

\bea \label{eq:oscillation}
G^{(2)}_{C,C/D}(t ,t') &\propto& \left| \int d \om{-} g(\om{-})  \osc ({\omega_- t_-}/{2})  \right|^2. \hspace{5mm}
\eea

\begin{figure}[t]
\centering
\includegraphics[width=0.95\columnwidth]{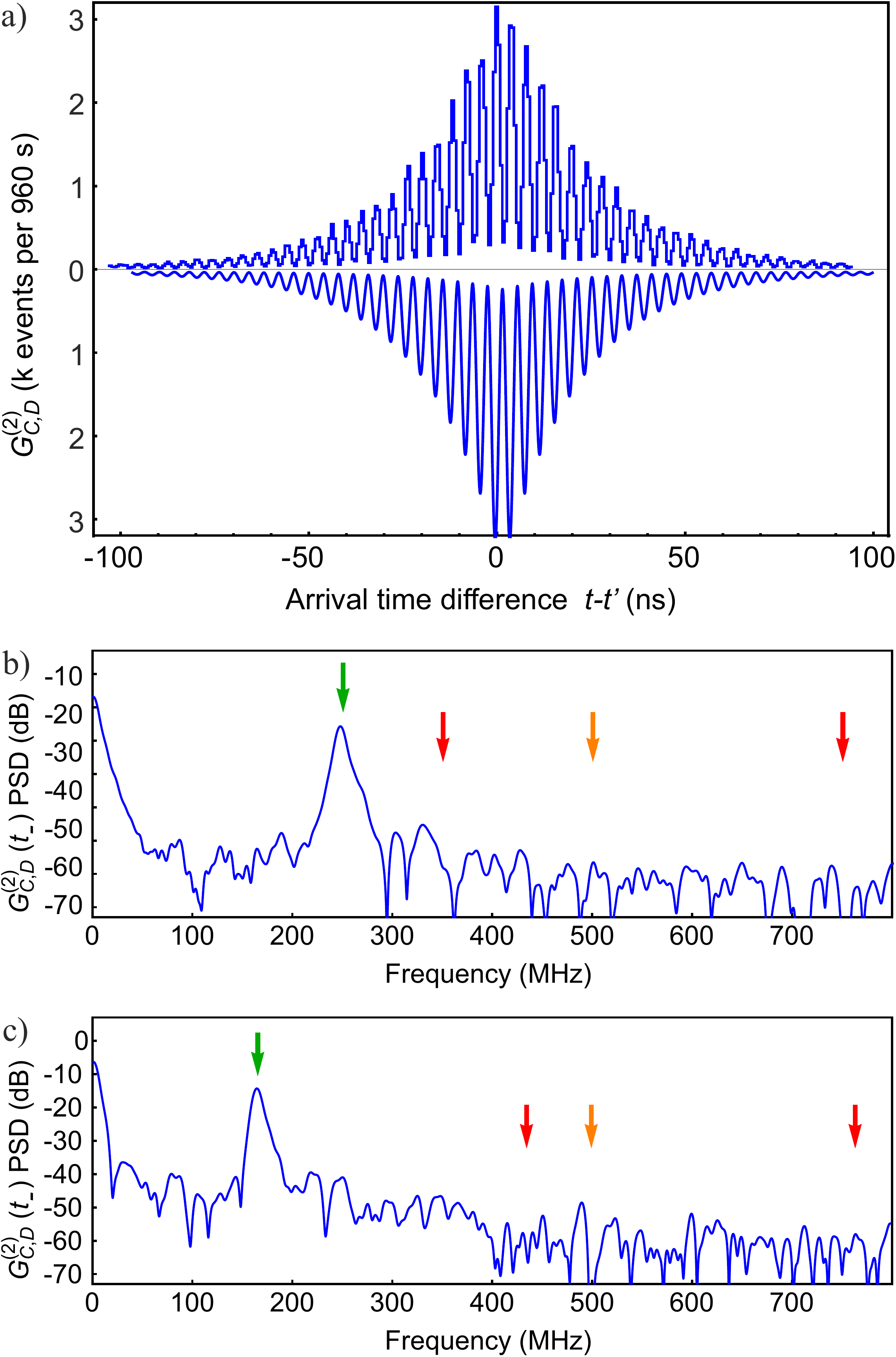} 

\caption{Observed and predicted $G^{(2)}_{C,D}$ cross correlation and spectral analysis. a) (upper curve) Histogram of recorded arrival time differences with \SI{625}{\pico\second} time bins, for $\omega^0_-=2\pi \times\SI{250}{\mega\hertz}$ . (lower curve)  Predicted $G^{(2)}_{C,D}$ with factors chosen manually such that the amplitude and visibility match the experimental results.  b) and c) Power spectral density (PSD) of $G^{(2)}_{C,D}(t_-)$ computed from observed histogram for $\omega^0_- =2\pi \times \SI{250}{\mega\hertz}$ and $2\pi \times \SI{165}{\mega\hertz}$ respectively. The expected beat-notes are clearly seen at \SI{250}{\mega\hertz} and \SI{165}{\mega\hertz} (green arrows). Locations at which spectral contamination might be expected are indicated with red and orange arrows. The contamination is at least \SI{25}{\decibel} below the power level of the desired beat note.   
}
\label{fig:HOMG2SignalsAndTransform}
\end{figure}

\autoref{fig:HOMG2SignalsAndTransform}a) shows the observed and predicted $G^{(2)}_{C,D}$ for a frequency difference of $\omega^0_- = 2 \pi \times \SI{250}{\mega\hertz}$ between the signal and idler modes. In accordance with the theory, the results show a clear oscillation with period \SI{4}{\nano\second}, the inverse of \SI{250}{\mega\hertz}. The visibility of the interference is 82$\%$, which is greater than the classical limit of 50$\%$ \cite{GhoshPRL1987}, attests to the fact that the interference was produced by non-classical states. The predicted $G^{(2)}_{C,D}$, calculated with $\gamma =2\pi\times\SI{7.6}{\mega\hertz}$, agrees with the observed fringe period and also with the decay rate of the exponential envelope. \autoref{eq:JSA} implies a FWHM bandwidth of $\gamma \sqrt{\sqrt{2}-1} = 2 \pi \times \SI{4.9}{\mega\hertz}$ for signal, idler, and difference frequency \cite{OBenson3}. The AHC measurement was repeated for a signal-idler frequency difference of \SI{165}{\mega\hertz}, and thus an oscillation period $ \approx \SI{6}{\nano\second}$ in  $G^{(2)}_{C,D}$.  

\begin{figure}[t]
\centering
\includegraphics[scale=0.27]{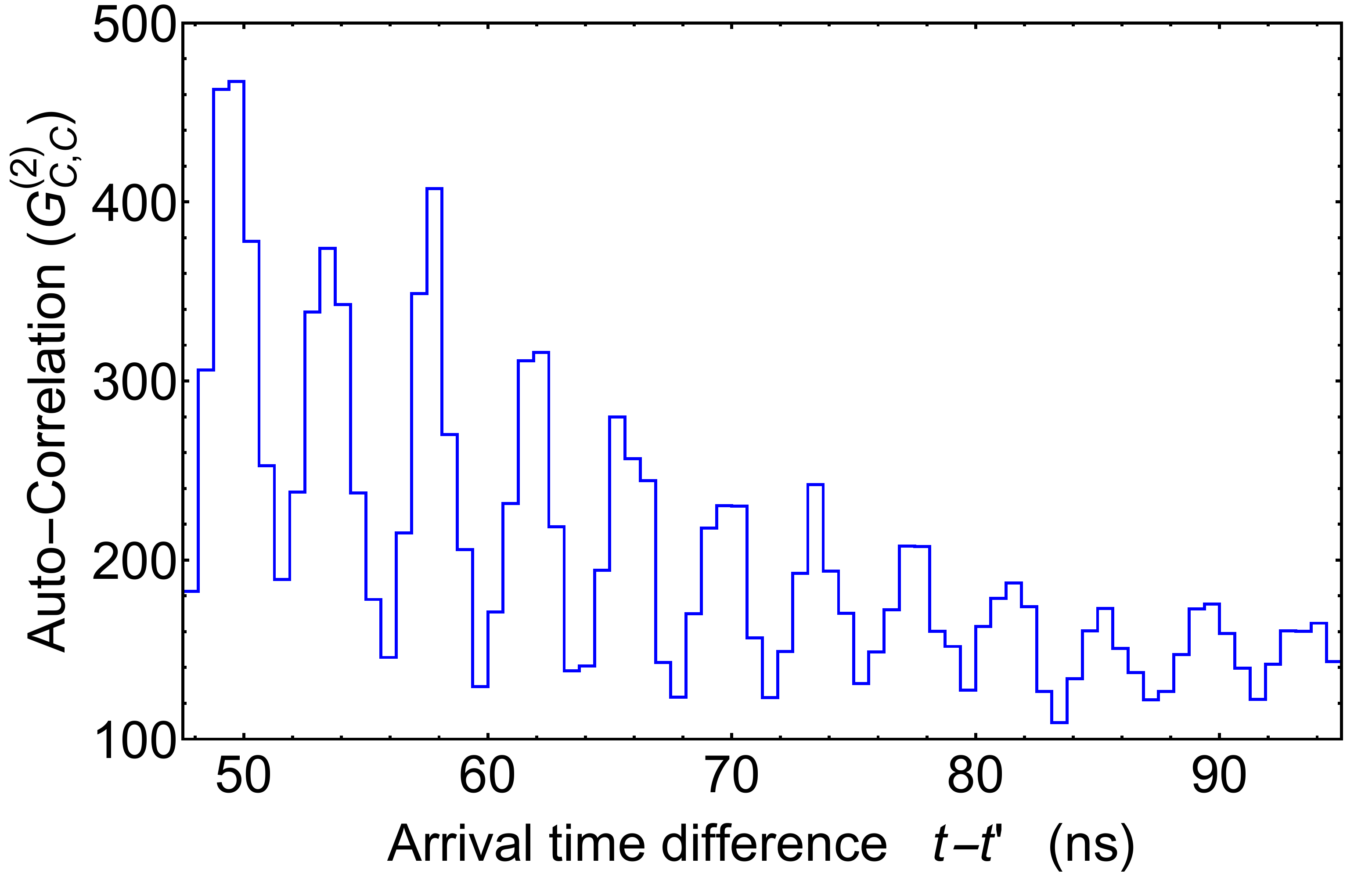} 
\caption{Temporal GM effect. Graph shows the un-normalised auto-correlation detected in detector $C$ ($G^{(2)}_{C,C}$) after the BS. The oscillations corresponding to the inverse frequency difference for $\omega^0_- = 2 \pi \times \SI{250}{\mega\hertz}$. }
\label{fig:GOMdata}
\end{figure}


The power spectra of the observed $G^{(2)}_{C,D}$ for $\omega^0_-/2\pi$ of \SI{250}{\mega\hertz} and \SI{165}{\mega\hertz} are shown in \autoref{fig:HOMG2SignalsAndTransform}b) and \ref{fig:HOMG2SignalsAndTransform}c) respectively. Each shows a peak at dc and a Lorentzian peak at the corresponding $\omega^0_-$, of the same width and center frequency as the Lorentzian in the JSI computed from \autoref{eq:JSA} \cite{Note1}. The resolution of these spectra is inversely proportional to the range of $t-t'$, which can extend to the full acquisition time. In practice, the resolution is much finer than any spectral feature. The spectral range, i.e. the largest observable difference frequency, is set by the  time resolution of the detection. With few-ps detection \cite{KorzhNP2020}, a difference-frequency range of $\sim \SI{100}{\giga\hertz}$ can be achieved \cite{Note1}. 



Inefficient extinction of neighbouring CE-SPDC modes by the FP filter would manifest as additional signals in the PSD  besides one at the expected $\om{-}^0$. In \autoref{fig:HOMG2SignalsAndTransform}b), peaks at \SI{750}{\mega\hertz} and/or \SI{350}{\mega\hertz} (aliased down from \SI{1250}{\mega\hertz} due to the sampling rate of the FPGA) would indicate leakage of mode pairs with $\omega_-/2\pi = \SI{250}{\mega\hertz} \pm 2\,\mathrm{FSR}$ (red arrows). A peak at \SI{500}{\mega\hertz} (the FSR for both signal and idler) might also be expected, but is not seen (orange arrow). Similarly, leakage would produce peaks at \SI{765}{\mega\hertz} and \SI{435}{\mega\hertz} (red arrows), and again \SI{500}{\mega\hertz} (orange arrow) in \autoref{fig:HOMG2SignalsAndTransform}c). We thus conclude that the FP cavity filtering succeeds in blocking contributions from neighbouring CE-SPDC modes and that the combined CE-SPDC and filter system emits on one pair of CE-SPDC cavity modes.

Measurement of $G^{(2)}_{C,C}$, showing the temporal GM effect, is shown in \autoref{fig:GOMdata} for $\omega^0_- = 2 \pi \times \SI{250}{\mega\hertz}$.  Typically SPDC $G^{(2)}$ autocorrelations require a BS and two detectors, in order to record photon pairs that arrive spaced by less than a detector's dead time, here $\approx \SI{40}{\nano\second}$. For narrow-band photon pairs, however, it is possible to acquire the autocorrelation with just one detector, as we do here.  The predicted oscillation with a period of \SI{4}{\nano\second} is clearly observed.  The temporal GM effect thus offers a simple way to characterize relative frequencies with a single detector. Frequency-shifting of one input photon through nonlinear-optical frequency conversion \cite{MaringO2018}, would allow AHC to measure spectra with difference frequencies outside the the detection electronics' bandwidth.

We have demonstrated a new technique to quantify the frequency correlations of narrow-band photon pairs and applied it to measure the spectral content of a filtered, cavity-enhanced parametric downconversion source at the Rb D$_1$ line with $\approx \SI{5}{\mega\hertz}$ optical bandwidth.  
By interfering the photons on a BS and {performing Fourier analysis on the} temporal auto- and cross-correlations, AHC directly measures the beat-note spectrum with the spectral resolution limited only by the acquisition time.  The technique is simple to implement with one detector via the temporal Ghosh-Mandel effect, or with two detectors via the non-degenerate Hong-Ou-Mandel effect. It is well suited to characterize narrow-band photon sources  for interaction with atoms and ions, which typically require bandwidths below  \SI{10}{\mega\hertz}. The technique may be especially valuable in quantum networking, computing and simulation with mixed photon-atom systems.

We thank Dr. Robert Sewell, Samyobrata Mukherjee and Nat{\'a}lia Alves for  feedback on the manuscript. This project was supported by  H2020 Future and Emerging Technologies Quantum Technologies Flagship project  QRANGE (Grant Agreement No.  820405); Spanish MINECO projects OCARINA (Grant No. PGC2018-097056-B-I00), Q-CLOCKS (Grant No. PCI2018-092973), and the Severo Ochoa program (Grant No. SEV-2015-0522); 
Generalitat de Catalunya through the CERCA program; 
Ag\`{e}ncia de Gesti\'{o} d'Ajuts Universitaris i de Recerca Grant No. 2017-SGR-1354;  Secretaria d'Universitats i Recerca del Departament d'Empresa i Coneixement de la Generalitat de Catalunya, co-funded by the European Union Regional Development Fund within the ERDF Operational Program of Catalunya (project QuantumCat, ref. 001-P-001644); Fundaci\'{o} Privada Cellex; Fundaci\'{o} Mir-Puig;
European Union's H2020 Marie Sk{\l{odowska-Curie Actions (665884). National Science Centre, Poland (2018/28/T/ST2/00275 and 2019/34/E/ST2/00440). 

%
%

\bibliographystyle{../biblio/apsrev4-1no-url}
\bibliography{../biblio/NDHOMBib210108}

\pagebreak
\widetext
  
\begin{center}
\textbf{\large Supplementary Material for \\ Autoheterodyne Characterization of Narrow-Band Photon Pairs}
\end{center}
\setcounter{equation}{0}
\setcounter{figure}{0}
\setcounter{table}{0}
\setcounter{page}{1}

\makeatletter
\renewcommand{\thepage}{S\arabic{page}} 
\renewcommand{\thesection}{S\arabic{section}}  
\renewcommand{\thetable}{S\arabic{table}}  
\renewcommand{\thefigure}{S\arabic{figure}}
\renewcommand{\theequation}{S\arabic{equation}} 

\newcommand{\sumdif}[1]{\mathring{#1}}
\newcommand{\fsumdif}{\sumdif{f}}
\newcommand{\Gsumdif}{\sumdif{G}}
\newcommand{\jta}{\tilde{f}}

\section{Preliminaries and definitions} 

\subsection{Definitions} 

As in the manuscript, we consider a two-photon, two-mode state $\ket{\psi}$, written in the general form
\begin{equation}
\label{eq:twophotonstate}
\ket{\psi} = \int d\omega_s d\omega_i \, f(\omega_s, \omega_i) \ha_s(\omega_s)  \ha_i(\omega_i) \ket{0},
\end{equation}
where $\omega_s$ and $\omega_i$ are, respectively, the signal and idler angular frequencies, $f(\omega_s, \omega_i)$ is the JSA, and $|f(\omega_s, \omega_i)|^2$ is the JSI.  In the time-domain, 
\be
\jta(t_s,t_i)  \propto  \int d\omega_s d\omega_i f(\omega_s,\omega_i)  e^{-\ii  \omega_s t_s} e^{-\ii  \omega_i t_i }
\ee
is the joint temporal amplitude (JTA) of the two-photon state.

In AHC, we introduce each mode of the two-photon state at one input face of a 50:50 beam splitter (BS) and look for correlations at the BS outputs $C,D$ of the form $G^{(2)}_{\kappa,\mu}(t,t') \equiv  |\langle 0 | \hEp_\kappa(t) \hEp_\mu(t')  | \psi \rangle|^2$,  $\kappa,\mu \in \{C,D\}$. Here the positive-frequency part of the fields at the detector are $\hEp_\kappa(t) \propto \int d\omega \ha_\kappa(\omega) \exp[{-i\omega t}]$ and $\ha_\kappa,$ is an annihilation operator. Due to the BS, the 
output fields are related to the signal-idler fields according to $\hEp_{C/D}(t)  = \frac{1}{\sqrt{2}} \left[ \hEp_s(t)  \pm \hEp_i(t) \right]$.

\subsection{Exchange antisymmetry or symmetry in the autoheterodyne measurement}

The nondegenerate HOM effect and the temporal GM effect can be understood via exchange symmetries.  Considering first the nondegenerate HOM signal, i.e. $G^{(2)}_{C,D}$, we find
\bea
\label{eq:G2HOM}
G^{(2)}_{C,D}(t,t') &=&  \left| \langle 0 | \hEp_C(t) \hEp_D(t')  | \psi \rangle \right|^2
\nne \left| \int d\omega_s d\omega_i   f(\omega_s, \omega_i) 
\frac{1}{2} \langle 0 |  [\hEp_s(t)+\hEp_i(t)]  [\hEp_s(t')- \hEp_i(t')] \had_s(\omega_s) \had_i(\omega_i) | 0 \rangle \right|^2 
\nnpropto \left| \int d\omega_s d\omega_i f(\omega_s, \omega_i)                                    \frac{1}{2}\left[e^{-i(\om{i}t+\om{s}t')}- e^{-i(\om{s}t+\om{i}t')}\right]\right|^2 \hspace{6mm} 
\nne 
\left| \frac{1}{2} [ \jta(t',t)-\jta(t,t')  ] \right|^2
\nnequiv \left| \jta_A(t,t') \right|^2,
\eea
where $\jta_A(t,t') \equiv \frac{1}{2} [ \jta(t',t)-\jta(t,t') ]$ is the exchange-antisymmetric part of  the JTA.

We note that $G^{(2)}_{C,D}(t ,t')$ can also be expressed directly in terms of the exchange-antisymmetric part of the JSA, as follows:
\bea
\label{eq:G2HOMfreq}
G^{(2)}_{C,D}(t ,t') &\propto&
\left| \int d\omega_s d\omega_i f(\omega_s,\omega_i)   \frac{1}{2} \left( e^{-\ii  (\omega_s t' + \omega_i t) } - e^{-\ii ( \omega_s t + \omega_i t' )} \right) \right|^2 
\nne \left|\int d\omega_s d\omega_i  e^{-\ii \omega_i t} e^{-\ii \omega_s t' }  \frac{1}{2}  f(\omega_s,\omega_i) - \int d\omega_s d\omega_i  e^{-\ii \omega_s t} e^{-\ii \omega_i t' }  \frac{1}{2}  f(\omega_s,\omega_i) 
 \right|^2 
\nne \left| \int d\omega' d\omega  e^{-\ii \omega t} e^{-\ii \omega' t' }  \frac{1}{2}  f(\omega',\omega)- \int d\omega d\omega' e^{-\ii \omega t} e^{-\ii \omega' t' }  \frac{1}{2}  f(\omega,\omega') 
 \right|^2 
\nne \left| \int d\omega d\omega' e^{-\ii \omega t} e^{-i \omega' t' }  \frac{1}{2} \left[  f(\omega',\omega) - f(\omega,\omega')  \right] \right|^2 
\nnequiv \left| \int d\omega d\omega' e^{-\ii \omega t} e^{-i \omega' t' }    f_A(\omega,\omega') \right|^2,
\eea
where $f_A(\omega,\omega') \equiv  \frac{1}{2} \left[  f(\omega',\omega) - f(\omega,\omega') \right]$ is the exchange-antisymmetric part of the JSA.   From \autoref{eq:G2HOM} and \autoref{eq:G2HOMfreq}, we see that $G^{(2)}_{C,D}(t ,t')$ is the square magnitude of the antisymmetric part of the JTA, or equivalently the 2D Fourier transform of the antisymmetric part of JSA.

In the same way, the temporal GM signals, i.e. $G^{(2)}_{C,C}$ and  $G^{(2)}_{D,D}$, are related to the exchange-symmetric part of the JSA: 

\bea
G^{(2)}_{\kappa,\kappa}(t ,t')  &=& \left| \int d\omega_s d\omega_i   f(\omega_s, \omega_i) 
\frac{1}{\sqrt{2}} \langle 0 |  \hEp_\kappa(t)\hEp_\kappa(t') \had_s(\omega_s) \had_i(\omega_i) | 0 \rangle \right|^2 \label{} 
\eea
where $\kappa \in \{C,D\}$. The factor $\frac{1}{\sqrt{2}}$, which does not appear in \autoref{eq:G2HOM}, is introduced to correctly predict the number of two-photon events detected when both photons leave the beam splitter by the same output. This is related to the normalization of the two-photon state $\bra{0}\hEp_\kappa(t)\hEp_\kappa(t') \propto \bra{0}\ha_\kappa\ha_\kappa = \sqrt{2}\bra{2}$.
\bea
\label{eq:G2GM}
G^{(2)}_{\kappa,\kappa}(t ,t')  &=& \left| \int d\omega_s d\omega_i   f(\omega_s, \omega_i) 
\frac{1}{2\sqrt{2}} \langle 0 |  [\hEp_s(t)+\hEp_i(t)]  [\hEp_s(t')+ \hEp_i(t')] \had_s(\omega_s) \had_i(\omega_i) | 0 \rangle \right|^2  \hspace{6mm}
\nnpropto 
\left| \int d\omega_s d\omega_i f(\omega_s,\omega_i)   \frac{1}{2\sqrt{2}} \left(  e^{-\ii ( \omega_s t + \omega_i t' )} +e^{-\ii  (\omega_s t' + \omega_i t) }  \right) \right|^2 
\nne 
\frac{1}{2}\left| \frac{1}{2} [ \jta(t,t') + \jta(t',t) ] \right|^2
\nnequiv \frac{1}{2}\left| \jta_S(t,t') \right|^2,
\eea
where $\jta_S(t,t') \equiv \frac{1}{2} [\jta(t,t') + \jta(t',t) ]$ is the exchange-symmetric part of the JTA.  Also  $G^{(2)}_{C,C}(t ,t')  = G^{(2)}_{D,D}(t ,t') \propto \frac{1}{2} \left| \int d\omega d\omega' e^{-\ii \omega t} e^{-i \omega' t' }  f_S(\omega,\omega') \right|^2$,  $f_S(\omega,\omega') \equiv  \frac{1}{2} \left[  f(\omega,\omega') +  f(\omega',\omega)  \right]$, in analogy to \autoref{eq:G2HOMfreq}.

\subsection{Sum and difference co-ordinates}

It is convenient to work in the sum and difference co-ordinates $\omega_\pm \equiv \omega_s \pm \omega_i$ and $t_\pm \equiv t \pm t'$ as the results simplify to the manuscript's Equation (2) when expressed in these transformed co-ordinates. The JSA in these co-ordinates can be written as
\bea
\label{eq:FSumDiff}
\fsumdif(\omega_+,\omega_-) & \equiv &  f\left( \frac{\omega_++\omega_-}{2} ,  \frac{\omega_+-\omega_-}{2}  \right). 
\eea
The antisymmetric part of this is 
\bea
\label{eq:FAsymSumDiff}
\fsumdif_A(\omega_+,\omega_-) & = & \frac{1}{2} [\fsumdif \left( \omega_+, -\omega_- \right)-  \fsumdif \left( \omega_+, \omega_- \right)   ] .
\eea
from which we obtain
\bea
\label{eq:G2HOMfreqSumDiff}
\Gsumdif^{(2)}_{C,D}(t_+ ,t_-) &\propto&
\left| \frac{1}{2}\int d\omega_+ d\omega_- e^{-\ii \omega_+ t_+/2} e^{-i \omega_- t_-/2}    \fsumdif_A(\omega_+,\omega_-) \right|^2
\eea
after noting that $\omega_+ t_+ + \omega_- t_- = 2(\omega_s t + \omega_i t')$.

\section{Interpretation of spectral features from AHC $G^{(2)}$}

As shown in the manuscript's Figure 3a, $\Gsumdif^{(2)}_{C,D}(t_+ ,t_-)$ shows oscillations that coincide with the beat-note between the most prominent signal and idler frequencies.  By looking at the PSD of $\Gsumdif^{(2)}_{C,D}(t_+ ,t_-)$,  we get information about the JSA and JSI, as illustrated in Figure 3b,c. By definition, the two-dimensional PSD of $G^{(2)}_{C,D}(t ,t')$ is proportional to the square magnitude of the Fourier transform of $G^{(2)}_{C,D}(t ,t')$
\bea
\label{eq:G2HOMPSA}
{\rm PSD}[ G^{(2)}_{C,D}(t ,t')](\omega,\omega')  &\propto&
\left| \int dt\, dt'\,  e^{\ii ( \omega t + \omega' t' )} G^{(2)}_{C,D}(t ,t')  \right|^2.
\eea
This can also be represented in the sum and difference coordinates
\bea
\label{eq:G2HOMPSAsumdif}
{\rm PSD}[ \Gsumdif^{(2)}_{C,D}(t_+ ,t_-)](\omega_+,\omega_-)  &\propto&
\left| \int dt_+ \, dt_-\,  e^{\ii ( \omega_+ t_+ + \omega_- t_- )} \Gsumdif^{(2)}_{C,D}(t_+ ,t_-)  \right|^2.
\eea
We now show how a spectral feature in the JSI translates to a spectral feature in ${\rm PSD}[ \Gsumdif^{(2)}_{C,D}]$. 

\subsection{Narrow-band pump scenario}

In the scenario of a narrow-band pump, as studied in the manuscript, the JSA and $G^{(2)}$ signals can be reduced to a function of a single coordinate.  Using the manuscript's Eq. (6), we express the JSA in the sum/difference coordinates as  
\bea
\label{eq:SumDiffNarrow}
\fsumdif_A(\omega_+,\omega_-) & \propto & \delta(\omega_+ - \omega_p) \left[  g(-\omega_-) - g(\omega_-) \right]
\eea
at which point we can perform trivially the integral over $d\omega_+$ that appears in  \autoref{eq:G2HOMfreqSumDiff}.  $\Gsumdif^{(2)}_{C,D}(t_+ ,t_-)$ is seen to be independent of $t_+$, 
\bea
\label{eq:G2HOMfreqNarrow}
\Gsumdif^{(2)}_{C,D}(t_+, t_-) &\propto&
\left| d\omega_-  e^{-i \omega_- t_-/2}   \left[ g(-\omega_-)- g(\omega_-) \right] \right|^2.
\eea
We henceforth omit the $t_+$ argument when discussing this narrow-band scenario.

\begin{figure}[t]
\centering
\includegraphics[width=0.65\textwidth]{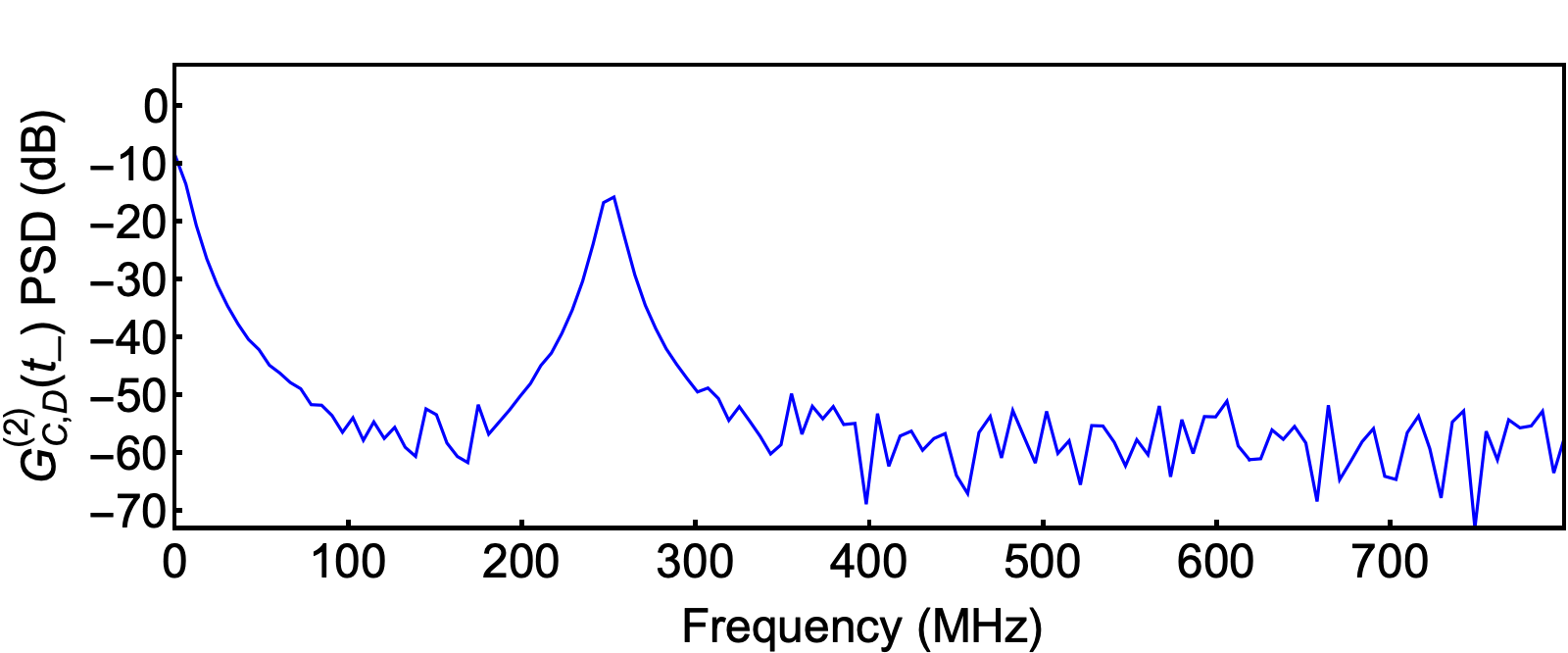}
\includegraphics[width=0.65\textwidth]{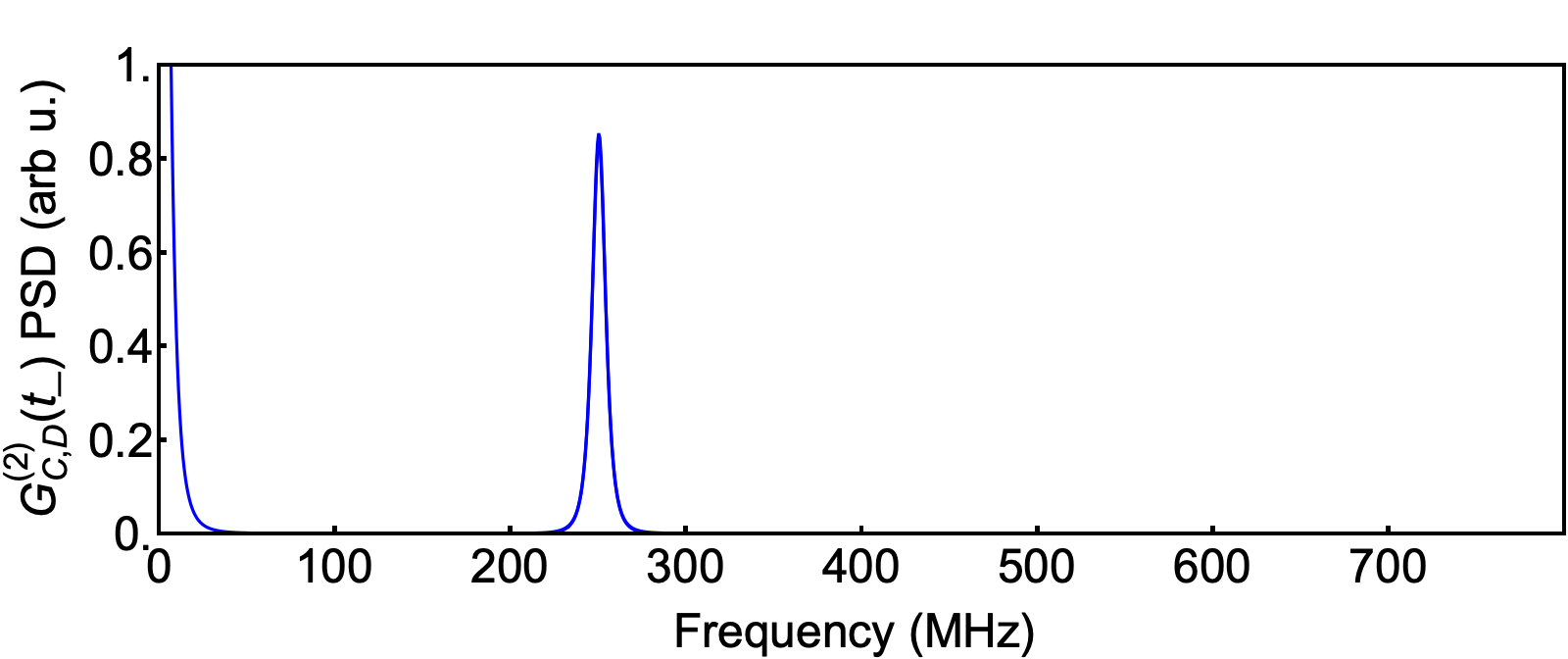}
\includegraphics[width=0.65\textwidth]{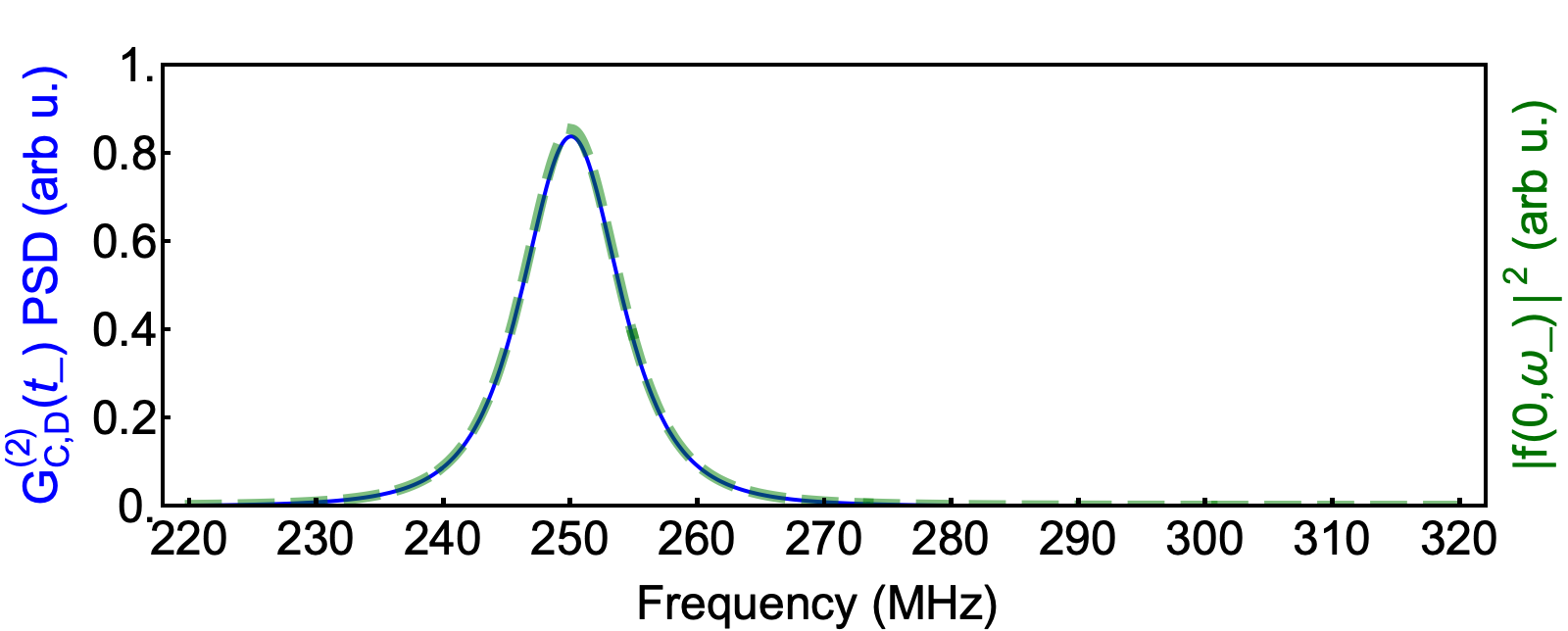}
\caption{Simulated spectra using \autoref{eq:G2HOMPSA}, a signal-idler beat-note of \SI{250}{\mega\hertz} and measured cavity linewidths from the article.  Top: PSD of $G^{(2)}_{C,D}(t_-)$ in log scale, (as in the manuscript).  A pseudo-random noise has been added to simulate the statistical noise of the $G^{(2)}_{C,D}$ acquisition. Middle: same as in the top graph, but on a linear scale.  Bottom: {close-up} of the feature at \SI{250}{\mega\hertz} (in blue), overlain with $|\fsumdif(\omega_+=0, \omega_-)|^2$ (in dashed green), showing the agreement of linewidth and line shape.    }
\label{fig:SimulatedSpectra}
\end{figure} 

\subsection{Illustration with two-photon state from the experiment}

Taking as an example the two-photon state described in the manuscript's Equation 6, we note that if $g(\omega_-)$ has a peak at some frequency, e.g. $ \omega^0_- \equiv \omega_s^0 - \omega_i^0$, then $g(-\omega_-) - g(\omega_-)$, will have a peak at {$ -\omega^0_-$ }and also at {$ \omega^0_-$}. These give a beat note in $\Gsumdif^{(2)}_{C,D}(t_-)$ at frequency $|\omega_i^0 - \omega_s^0|$. Note that the frequency doubling (beating of $ \omega_-^0$ with $- \omega_-^0$, a result of antisymmetrization) cancels the factor of $1/2$ due to the use of sum and difference coordinates.  We thus see that peaks in $g(\omega_-)$ will be represented as peaks in the PSD of  $\Gsumdif^{(2)}$ at the same frequencies.

We illustrate this, again using the manuscript's Eq. (6), and with spectral parameters taken from the experiment, in \autoref{fig:SimulatedSpectra}.  Also shown is the JSI $|\fsumdif(\omega_+=0, \omega_-)|^2$, which illustrates the fact that the line width in the JSI and in the PSD of $\Gsumdif^{(2)}_{C,D}(t_-)$ are the same. The width of the {peak} in the $G^{(2)}$ PSD shown in the manuscript's Fig. 3 is the spectral width, along the $\omega_-$ direction, of the two-photon JSI.  As described in \ref{sec: freq resolution}, broadening from the measurement technique is negligible.

\subsection{Further observations}

We note that $G^{(2)}$, like the JSI, scales quadratically with $f$ (the JSA), and thus that ${\rm PSD}[ G^{(2)}]$ scales as the fourth power of $f$.  This nonlinear relationship between the JSI and ${\rm PSD}[ G^{(2)}]$ leads to a number of effects. First and most simply, the height of a peak in ${\rm PSD}[ G^{(2)}]$ is proportional to the square of the height of the corresponding peak in the JSI.  Second, ${\rm PSD}[ G^{(2)}]$ has many of the properties of a convolution (in suitable coordinates) of the JSI with itself. This, for example, explains the origin of the peak in ${\rm PSD}[ G^{(2)}]$ around zero frequency. To continue with the example just described, if $|f_A(\omega_s,\omega_i)|^2$  has peaks at $ \omega^0_-$ and $- \omega^0_-$, then the autoconvolution will (again taking into account the factor of $1/2$ from the change of coordinates) have peaks at zero, $ \omega^0_-$ and $-\omega^0_-$,  as seen in the manuscript's Figure 3b,c and \autoref{fig:SimulatedSpectra}.  A more complex JSI, e.g. with multiple difference-frequency peaks, would lead to cross-terms at frequencies that are sums and differences of the frequencies represented in the JSI.

\subsection{Temporal Ghosh-Mandel signals}

The signals for the temporal Ghosh-Mandel effect ($G^{(2)}_{CC}$ and $G^{(2)}_{DD}$) are analogously related to $\jta_S(t,t')$, and result in a very similar spectrum. Due to conservation of photon number,  $G^{(2)}_{CC}+G^{(2)}_{DD}$ and $G^{(2)}_{CD}$ must sum to give $G^{(2)}_{AB}$, the output of the SPDC cavity before HOM mixing on the beam splitter.  $G^{(2)}_{AB}$ does not exhibit a beat-note; it's spectrum is limited to the two-photon bandwidth. The oscillating components of $G^{(2)}_{CC}+G^{(2)}_{DD}$ and of $G^{(2)}_{CD}$ are thus equal and opposite, leading to the same peaks in their ${\rm PSD}[G^{(2)}]$  representation.

\section{Application of AHC to pulsed CE-SPDC and heralded pure state generation}

One major motivation for studying two-photon states is their potential use in quantum networking, in which photons from different sources are interfered for purposes such as entanglement swapping.  These applications typically require two-photon states in which the frequency entanglement is small, such that one photon is left in a pure or nearly-pure spectral state when the other photon is traced over.  Such systems typically employ pulsed pump fields, which have the effect of broadening the JSA along the $\omega_+$ dimension, thereby reducing the signal-idler frequency correlations. Here we study this scenario and show how AHC can be applied to it. 

We write the pump field $E_p(t)$ in terms of its Fourier amplitudes $\alpha(\omega_p)$
\begin{equation}
E_p(t)= \int d\omega_p\alpha(\omega_p)e^{-i \omega_p t}.
\end{equation}
The two-photon state {from the filtered CE-SPDC system} is as in \autoref{eq:twophotonstate} above, with 
\bea
\label{eq:JSA}
f(\omega_s,\omega_i)&\propto& \int d\omega_p \delta(\omega_p-\omega_s - \omega_i) \alpha(\omega_p) \frac{1}{{\gamma}/{2}+\ii\:(\omega^0_{s}-\omega_s)}  \frac{1}{{\gamma}/{2}+\ii\:(\omega^0_{i}-\omega_i)}.
\eea
Here $\omega^0_{s,i}$  and $\gamma$ are the center frequencies and linewidth of two cavity modes (compare to Equation 6 in the manuscript, which is the same except with a monochromatic pump so that $\alpha(\omega_p)$ is a delta function).  If we take $\alpha(\omega_p)$ to be a Gaussian of rms width $\sigma_p$, we have the JSA for a narrow-band SPDC source with a pulsed pump.  


\begin{figure}[t]
\begin{minipage}[c][1\width]{0.45\textwidth}
\centering
	\includegraphics[width=1\textwidth]{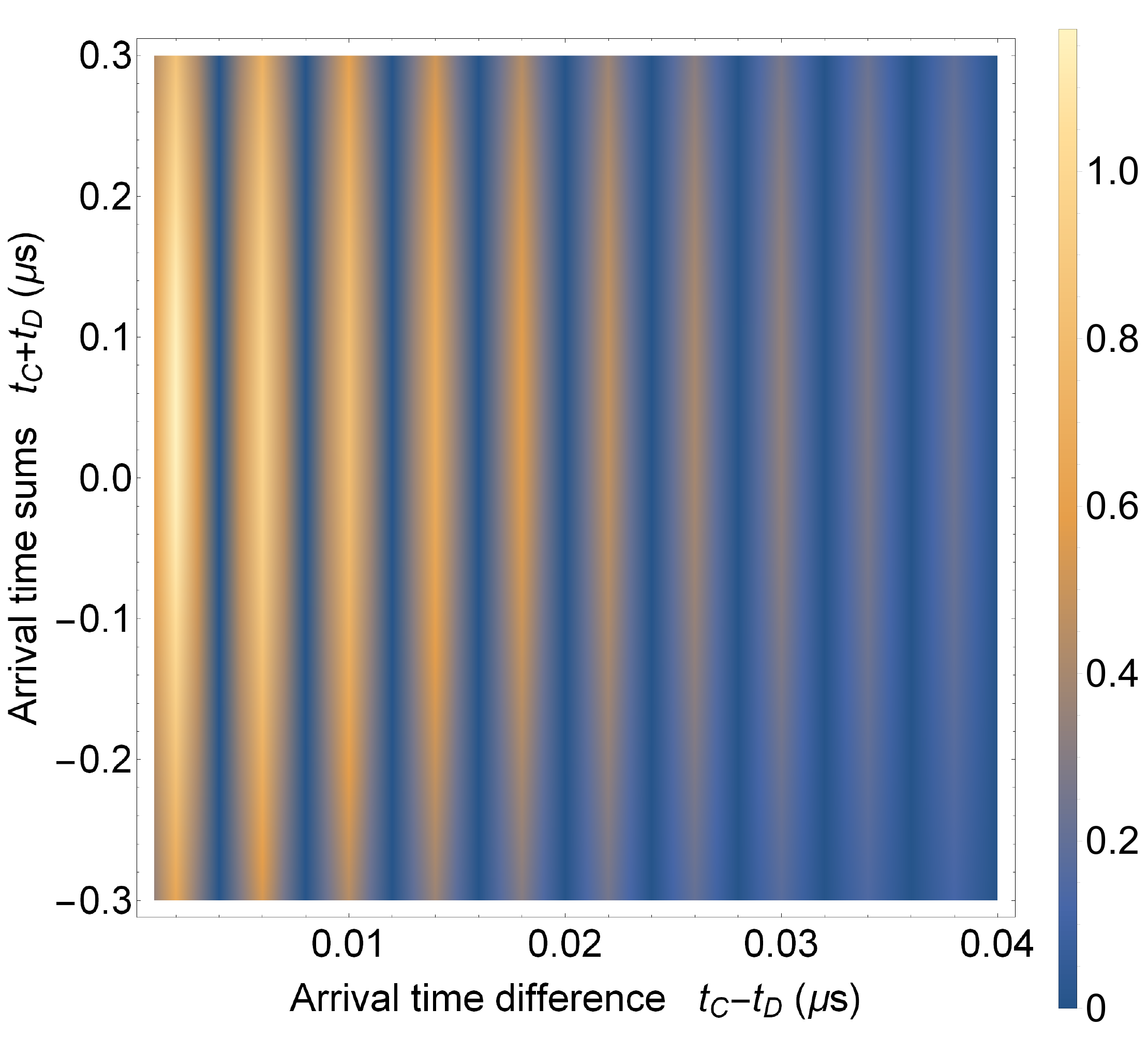}
	\subfloat{(a) $G^{(2)}_{C,D}$ for $\sigma_p =  2 \pi \times \SI{0.5}{\mega\hertz}$}
\end{minipage}
 \hfill 	
\begin{minipage}[c][1\width]{0.45\textwidth}
	\centering
	\includegraphics[width=1\textwidth]{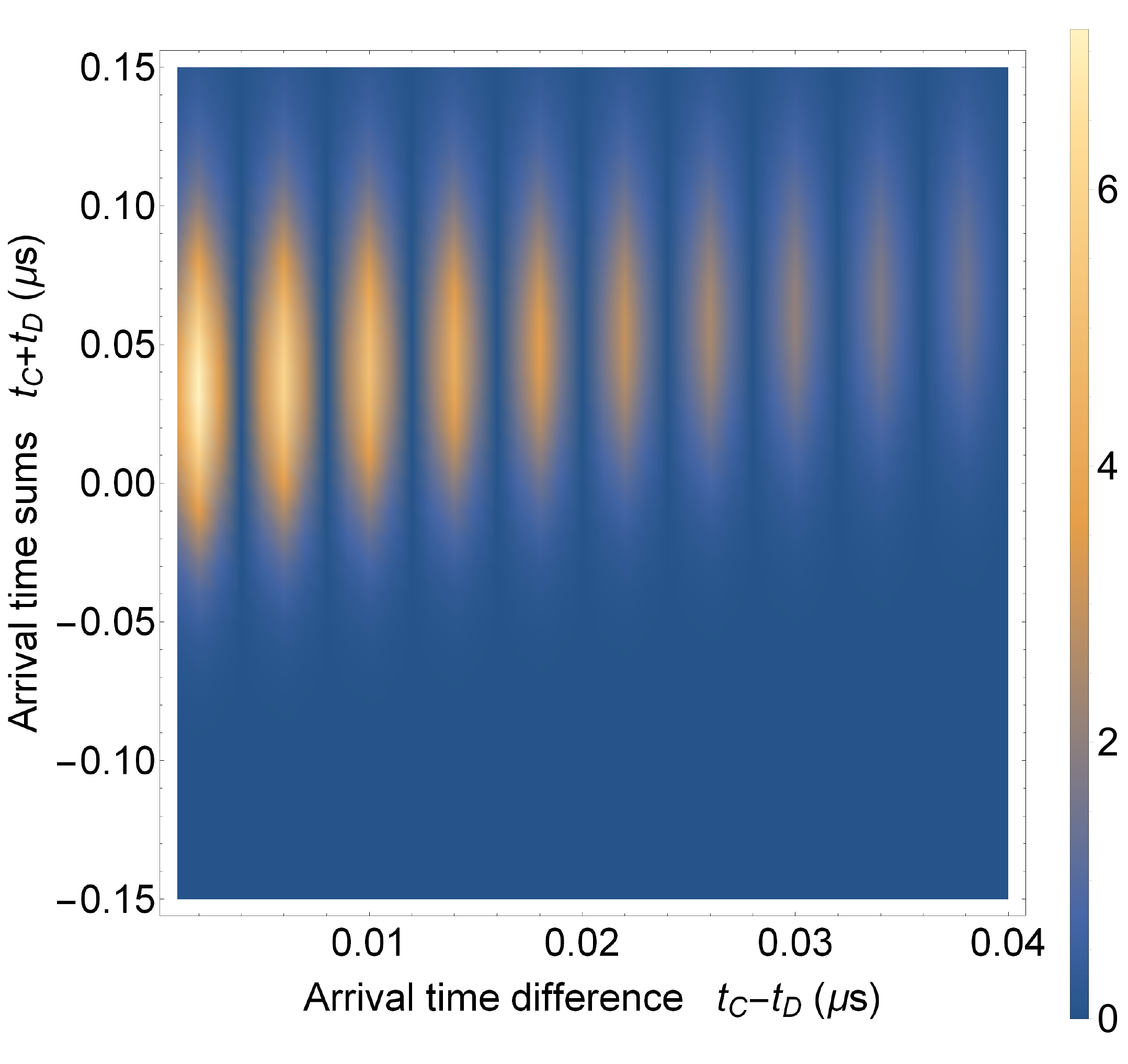}
	\subfloat{(b) $G^{(2)}_{C,D}$ for $\sigma_p =  2 \pi \times \SI{5}{\mega\hertz}$}
\end{minipage}
\caption{Modelled autoheterodyne coincidences $G^{(2)}_{C,D}$ for various pump bandwidths, for a CESPDC cavity bandwidth of \SI{7}{\mega\hertz} and $\om{-}^0=2 \pi  \times \SI{250 }{\mega\hertz}$. Oscillations in the coincidences along the $t_-$ axis correspond to the inverse of the frequency spacing. The drop in coincidences along the $t_-$ axis  gives the 2-photon correlation time or bandwidth. The Gaussian profile of coincidences in the $t_+$ axis has a FWHM which is the inverse of the pump bandwidth. The $t_+$ distribution is centred on $t_+ \approx \SI{0.05}{\micro\second}$, which is the group delay produced by the filter cavities. }
\label{fig:G2model}
\end{figure}

\begin{figure}[h!]
	\begin{minipage}[t]{0.4\textwidth}
	   \centering
	   \includegraphics[width=0.9\textwidth]{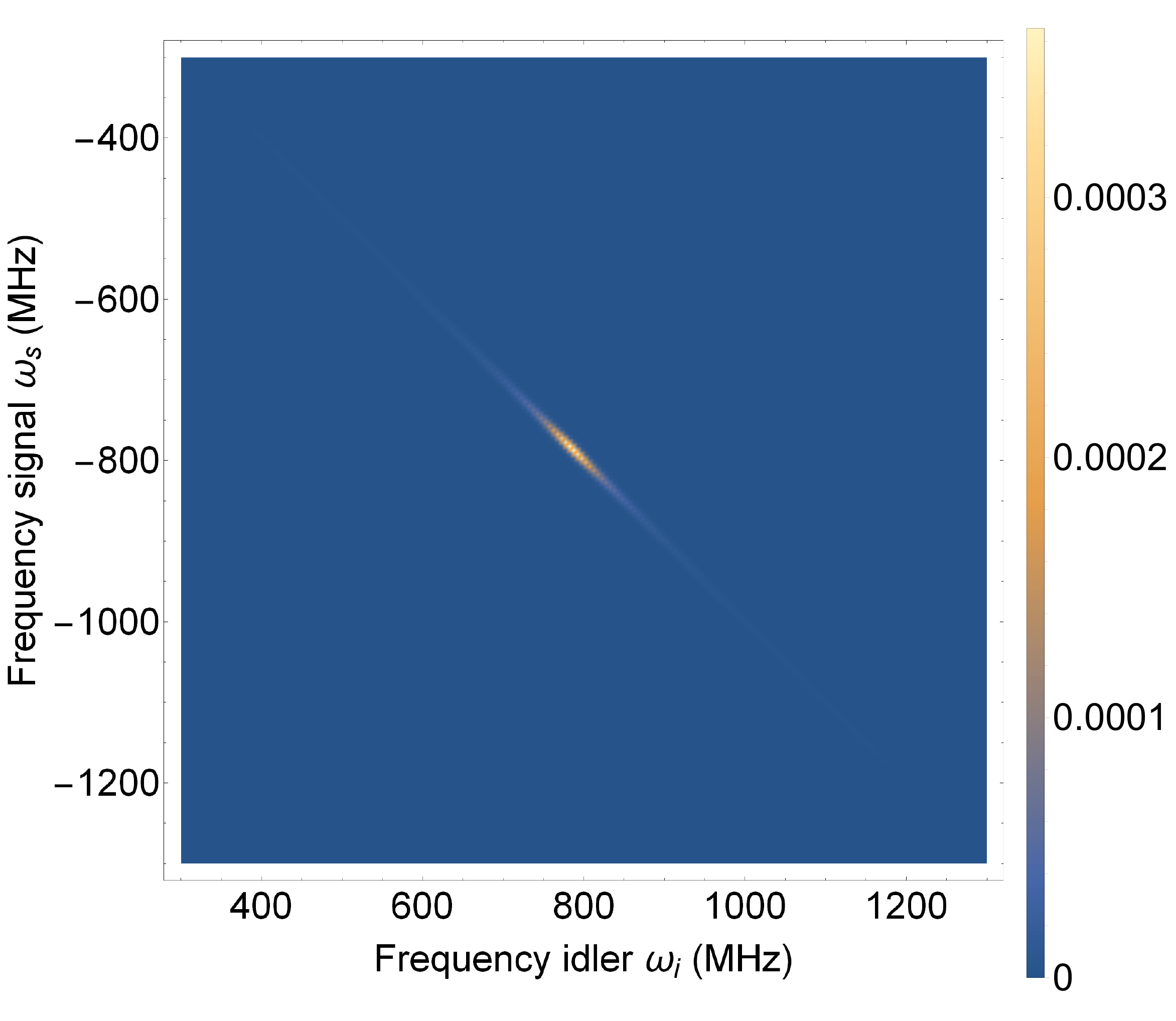}
	   \subfloat{(a) JSA magnitude $|f(\omega_s,\omega_i)|$ for $\sigma_p =  2 \pi \times \SI{0.5}{\mega\hertz}$}
	\end{minipage}
\hfill	
	\begin{minipage}[t]{0.4\textwidth}
	   \centering
	   \includegraphics[width=0.9\textwidth]{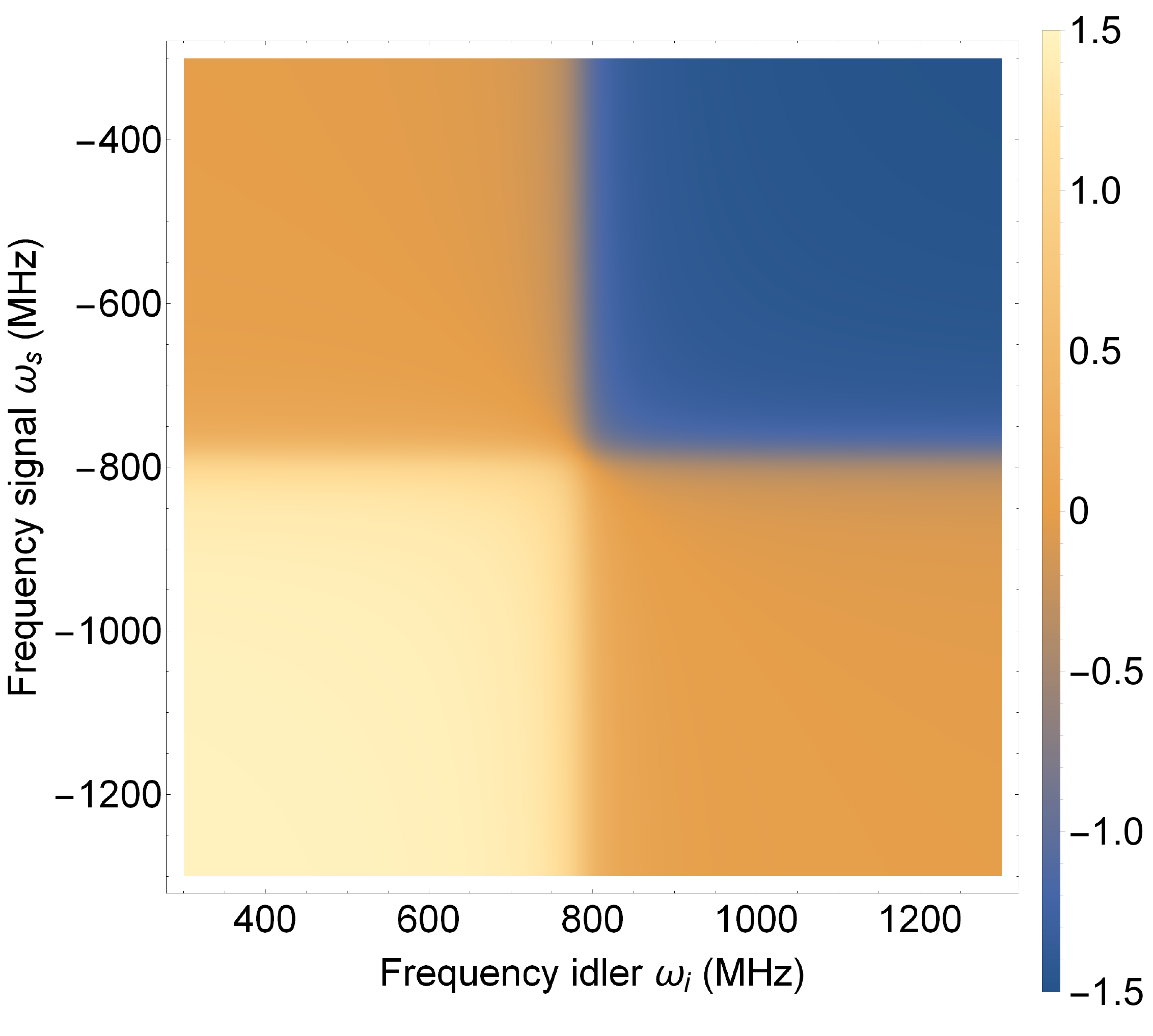}
	   \subfloat{(b) JSA phase $\arg[f(\omega_s,\omega_i)]$ for $\sigma_p =  2 \pi \times \SI{0.5}{\mega\hertz}$}
	\end{minipage}
	\begin{minipage}[t]{
	   0.4\textwidth}
	   \centering
	   \includegraphics[width=0.9\textwidth]{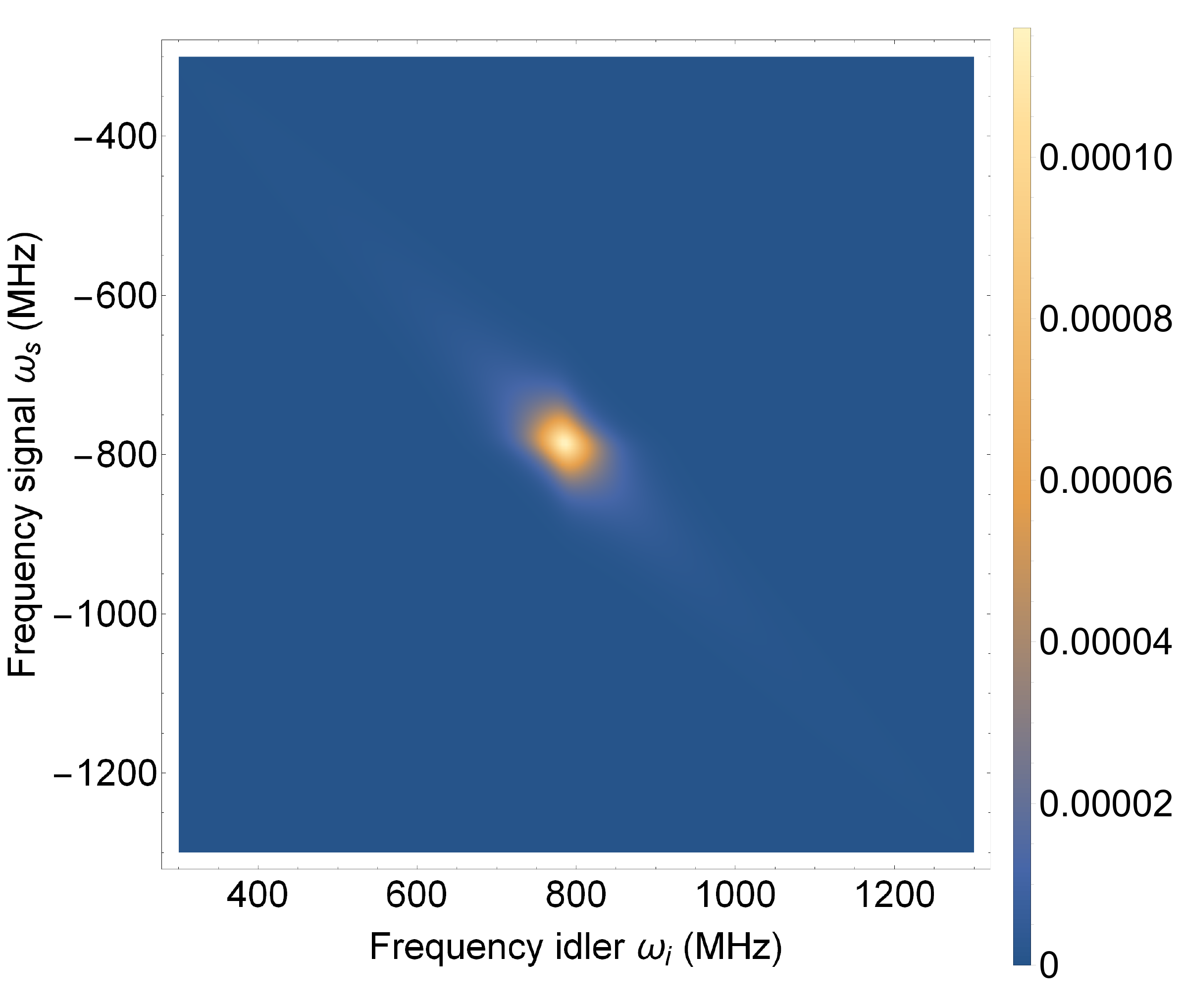}
	   \subfloat{(c) JSA magnitude $|f(\omega_s,\omega_i)|$ for $\sigma_p =  2 \pi \times \SI{5}{\mega\hertz}$}
	\end{minipage}
\hfill	
	\begin{minipage}[t]{
	   0.4\textwidth}
	   \centering
	   \includegraphics[width=0.9\textwidth]{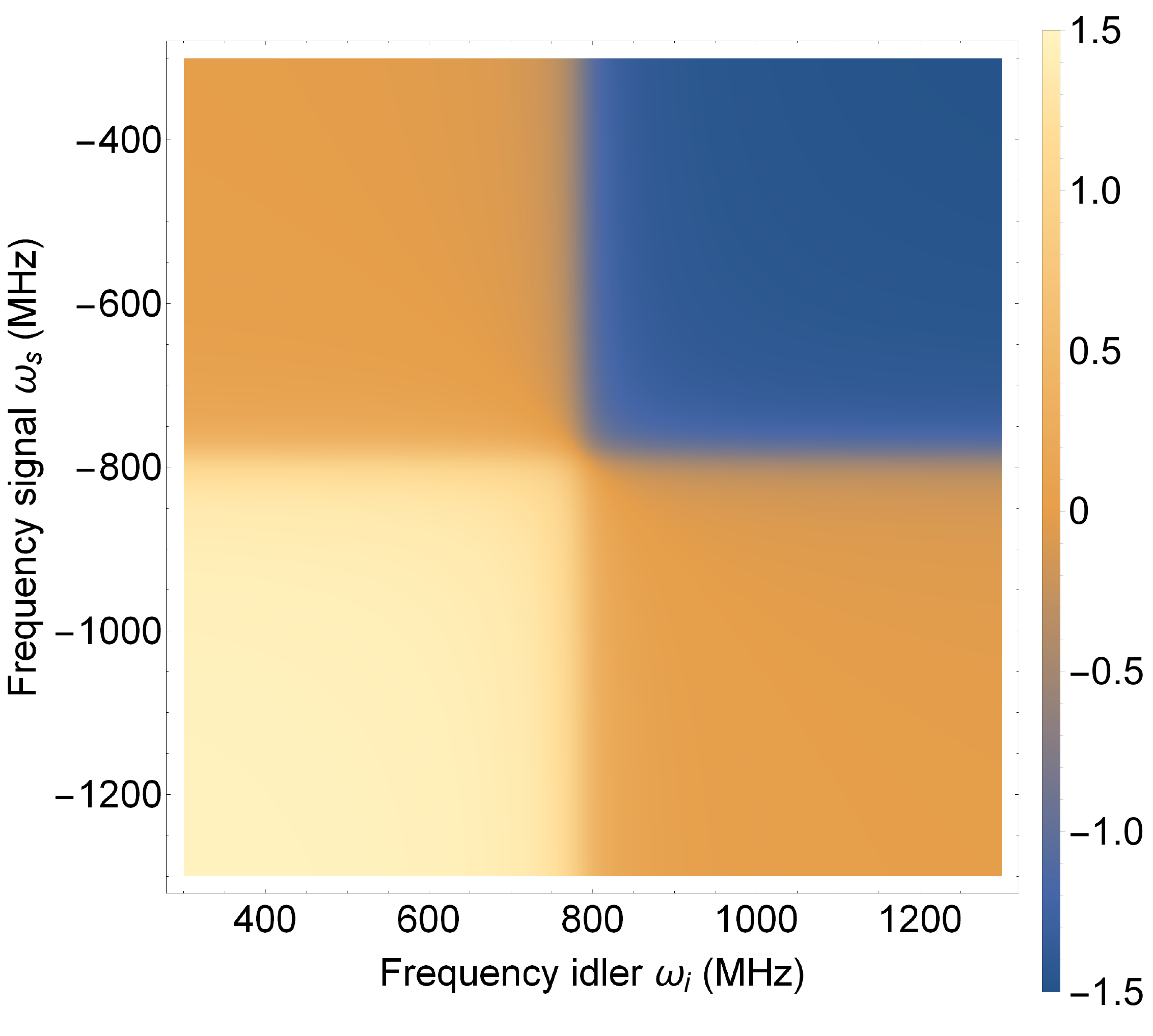}
	   \subfloat{(d) JSA phase $\arg[f(\omega_s,\omega_i)]$ for $\sigma_p =  2 \pi \times \SI{5}{\mega\hertz}$}
	\end{minipage}
	\begin{minipage}[t]{0.4\textwidth}
	   \centering
	   \includegraphics[width=0.9\textwidth]{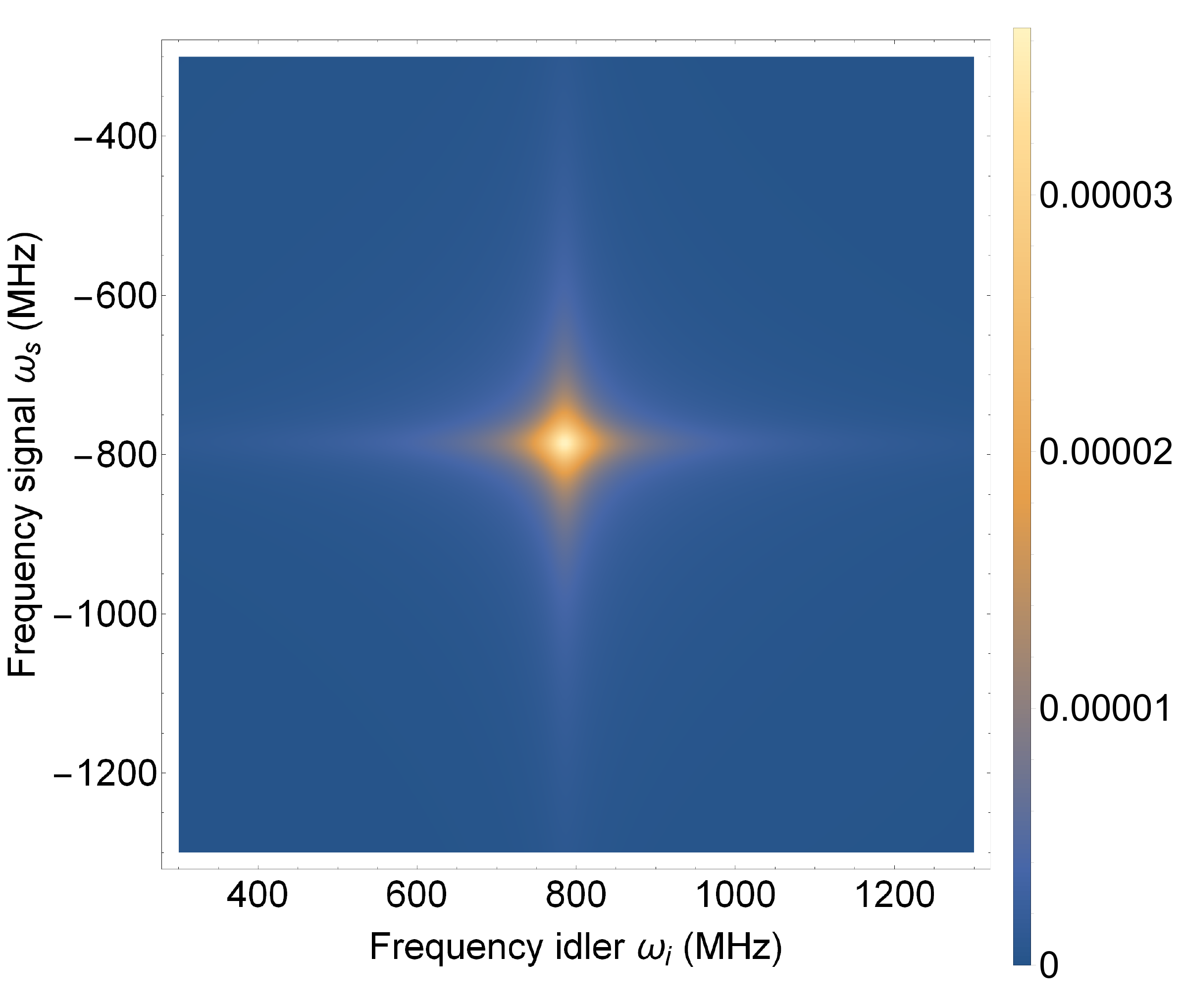}
	   \subfloat{(e) JSA magnitude $|f(\omega_s,\omega_i)|$ for $\sigma_p =  2 \pi \times \SI{50}{\mega\hertz}$}
	\end{minipage}
\hfill	
	\begin{minipage}[t]{
	   0.4\textwidth}
	   \centering
	   \includegraphics[width=0.9\textwidth]{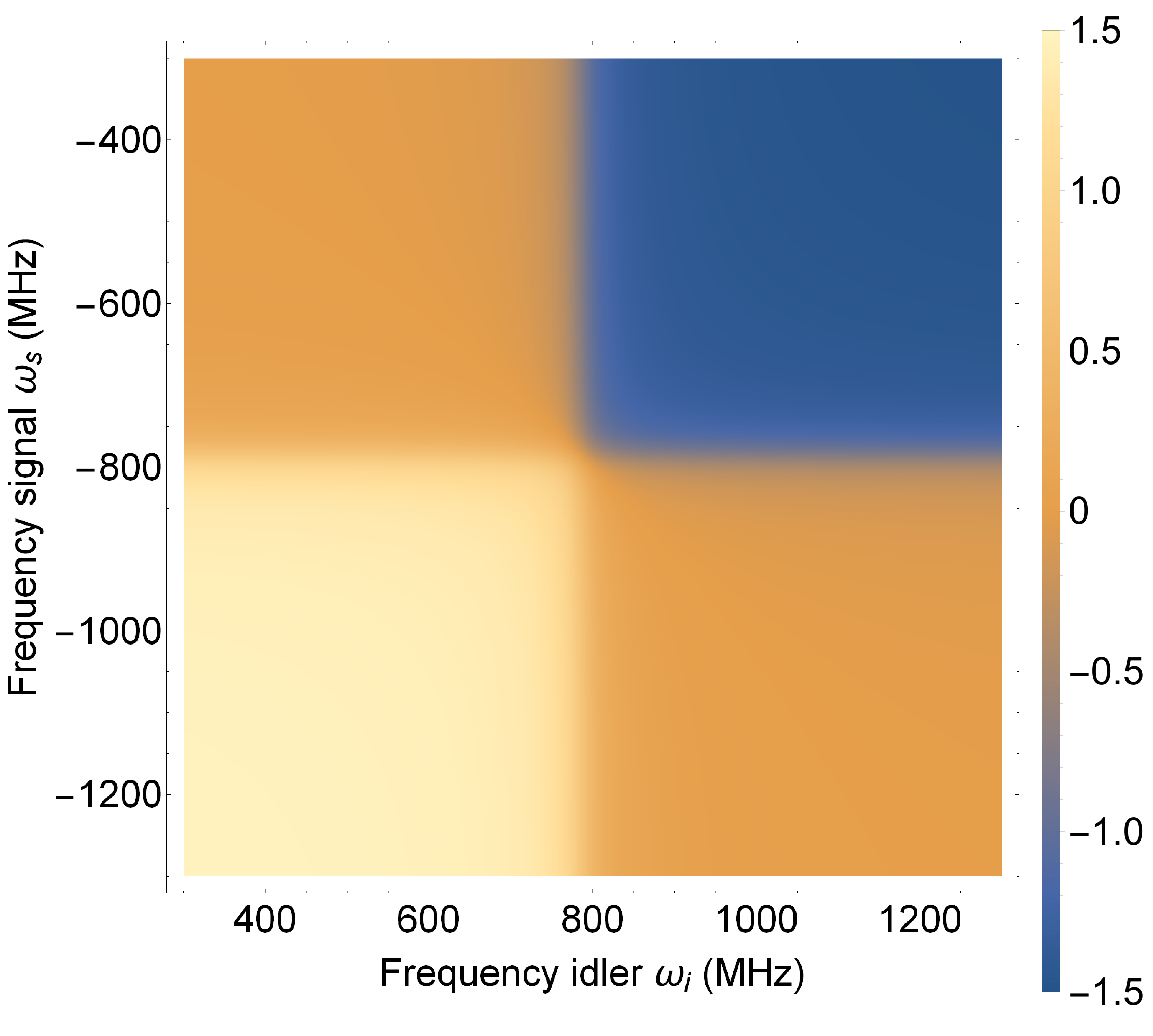}
	   \subfloat{(f) JSA phase $\arg[f(\omega_s,\omega_i)]$ for $\sigma_p =  2 \pi \times \SI{50}{\mega\hertz}$}
	\end{minipage}
\caption{JSA for various pump bandwidths inferred from autoheterodyne results. Plots (a) and (b) correspond to the JSA when the pump is narrow-band. In this scenario the photons are highly correlated with an entanglement entropy of 4.4. Plots (c) and (d) correspond to a pump bandwidth comparable to the 2-photon bandwidth and thus there is lesser entanglement with the down-converted photons. The entanglement entropy in this case is 1.8. Plots (e) and (f) give the JSA for a very broadband pump and the entanglement entropy here is 0.2, much closer to the limit of 0 for a pure state. }
\label{fig:JSA}
\end{figure}

We now show how AHC can be applied to a state generated in this way. The two-photon state in the sum and difference co-ordinates is,
\begin{equation}
\ket{\psi} = \int \frac{1}{2}d\omega_+ d\omega_- \, \fsumdif_A(\omega_+,\omega_-) \had_s\left(\frac{\omega_+ + \omega_-}{2}\right)  \had_i\left(\frac{\omega_+ - \omega_-}{2}\right) \ket{0}
\end{equation}
Performing the integral over $\omega_p$ in \autoref{eq:JSA} we find 
\be 
\fsumdif_A(\omega_+,\omega_-) \propto \alpha(\om{+}) \frac{1}{{\gamma}+\ii\:(2\omega^0_{s}-\omega_+ - \omega_-)} \frac{1}{{\gamma}+\ii\:(2\omega^0_{i}-\omega_+ + \omega_-)}.
\ee

Using this in the manuscript's Equation 2, we compute the AHC signals. The correlation function $G^{(2)}_{CD}(t, t')$ for various pump bandwidths is shown in  \autoref{fig:G2model}. The other AHC signals, $G^{(2)}_{CC}(t, t')$ and $G^{(2)}_{DD}(t, t')$ look similar, but the ``beating,'' i.e. oscillation with $t-t'$, is 180$^\circ$ out of phase.  These signals are directly observable by measuring the arrival times of the photons at detectors C and D (see manuscript Figure 1), and provide spectral information about the JSA.  As described in the manuscript's  Equation 2, the various $G^{(2)}$ functions can be expressed as the square magnitude of a two-dimensional Fourier transform (the usual complex Fourier transform along the $\omega_+$ axis and either a sine or cosine transform along the $\omega_-$ axis).

As with more established methods \cite{zielnicki2018jsi}, the AHC observations can be compared against the theoretical model, for validation and to estimate parameters such as $\gamma$, $\sigma_p$ and $\omega_s^0 - \omega_i^0$. From the model's JSA, with parameters found by measurement, the Schmidt number and entanglement entropy can be estimated using the methods described in \cite{BranczykNJP2010}. Examples are shown graphically in \autoref{fig:JSA}. The state purity and other figures of merit can then be computed from the model. 

%

\noindent

\section{Frequency resolution and range of the AHC technique}\label{sec: freq resolution}

Spectral characterization by the AHC technique is based on time-correlated photon counting, not on optical frequency discrimination, e.g. using a monochromator. For this reason, its spectral resolution is  not limited by the resolution of any optical instrument. It is, rather, limited by frequency resolution of the electronic systems used to record the photon detections. For the experiment reported here, this limit would be set by the frequency instability of the electronic clock used in the time-tagging module, and ultimately by the duration of the acquisition.  In practice, these resolution limits allow measurement of the finest spectral features of any practical SPDC source and contribute negligible broadening to the inferred spectra, e.g. those shown in the manuscript's Figure 3.

The spectral  range of the AHC technique is limited by the time resolution (or jitter) of the detectors and time-tagging electronics.  Although not the case for our equipment, this time resolution can be in the few-picosecond regime with modern detectors and electronics, giving a $\sim \SI{100}{\giga\hertz}$ beat-note bandwidth for the technique \cite{KorzhNP2020}.  If the frequency difference to be investigated is larger than this, nonlinear optical frequency conversion could be used to shift the frequency of one photon while preserving its nonclassical features \cite{MaringO2018}, and thereby bring the beat note and sidebands within the range of the technique. 

\section{On the possibility of spectral characterization by first-order interference}

{A classical beat-note measurement would combine two light fields of different frequencies into a single mode, for example with a beam splitter, and then detect the resulting oscillation of the intensity.  The average signal in such an experiment is described by the first order correlation function $G^{(1)}(t)\equiv \langle \psi | \hEm_C(t) \hEp_C(t)  | \psi \rangle$.

While this kind of measurement can produce an oscillating power or intensity when two coherent states of different frequencies are combined, the same is not true for two single-photon states.  This is easily demonstrated:   If the state is $\ket{\psi} = \int d\omega_s d\omega_i f(\omega_s, \omega_i) \had_s(\omega_s) \had_i(\omega_i) \ket{0}$, then 
\bea
G^{(1)}(t) &\equiv&  \langle \psi | \hEm_C(t) \hEp_C(t)  | \psi \rangle \nne
\label{eq:NoInterferenceA}
 \int d\omega_s d\omega_i  \int d\omega_s' d\omega_i' f^*(\omega_s, \omega_i)  f(\omega_s', \omega_i')  \nnt
\frac{1}{2} \langle 0 |\ha_s(\omega_s) \ha_i(\omega_i)  [\hEm_s(t)+\hEm_i(t)]  [\hEp_s(t)+\hEp_i(t)] \had_s(\omega_s') \had_i(\omega_i') | 0 \rangle \hspace{6mm}
\\ 
\label{eq:NoInterferenceB}
& = & 
 \int d\omega_s d\omega_i  \int d\omega_s' d\omega_i' f^*(\omega_s, \omega_i)  f(\omega_s', \omega_i')  \nnt
\frac{1}{2} \langle 0 |\ha_s(\omega_s) \ha_i(\omega_i)  [\hEm_s(t) \hEp_s(t)+\hEm_i(t)\hEp_i(t)] \had_s(\omega_s') \had_i(\omega_i') | 0 \rangle
\\ & = &
\label{eq:NoInterference}
\frac{1}{2} [ G^{(1)}_s(t) + G^{(1)}_i(t) ],
\eea
where  $G^{(1)}_s(t) =  \langle \psi | \hEm_s(t) \hEp_s(t)  | \psi \rangle$ and similarly  $G^{(1)}_i(t)$ are the average power of the individual beams.  We see from \autoref{eq:NoInterference} that there is no first-order beating or interference of these fields, as a consequence of their being in a twin Fock state.  This is evident in the vanishing of the cross-terms in going from \autoref{eq:NoInterferenceA}  to \autoref{eq:NoInterferenceB}, 
 e.g. 
\bea
 \langle 0 |\ha_s(\omega_s) \ha_i(\omega_i) \hEm_s(t) \hEp_i(t) \had_s(\omega_s') \had_i(\omega_i') | 0 \rangle = 0. 
\eea
For this reason, an ordinary, i.e.  $G^{(1)}(t)$, beat note measurement between photon pairs would provide no information at all about their relative frequencies.  In contrast, the coincidence measurements are described by a second-order correlation function, which does indeed provide information about the relative frequency of  the photons, as shown in the article. 
}


\end{document}